\documentclass[twocolumn,english,amsmath,amssymb,superscriptaddress,nofootinbib]{revtex4-1}
\usepackage{mathptmx}
\usepackage[T1]{fontenc}
\usepackage[latin9]{inputenc}
\usepackage{babel}
\usepackage{amsmath}
\usepackage{amssymb}
\usepackage{graphicx}
\usepackage{esint}
\usepackage[unicode=true,pdfusetitle,
 bookmarks=true,bookmarksnumbered=true,bookmarksopen=false,
 breaklinks=false,pdfborder={0 0 1},backref=false,colorlinks=false]
 {hyperref}
 \usepackage{color}

\newcommand{\cob}[1]{{\color{blue} #1}}

\makeatletter

\newcommand*\LyXThinSpace{\,\hspace{0pt}}

\makeatother

\begin{document}

\title{Collective Description of  Density Matrix of Identical Multi-level Atoms for Superradiance }

\author{Yuan Zhang}
\email{yzhuaudipc@163.com}

\address{School of Physics and Microelectronics, Zhengzhou University, Daxue Road 75, Zhengzhou 450052 China}
\address{Donostia International Physics Center, Paseo Manuel de Lardizabal 4, Donostia-San Sebastian (Gipuzkoa) 20018 Spain}
\address{Department of Physics and Astronomy, Aarhus University, Ny Munkegade 120, Aarhus C DK-8000 Denmark}

\begin{abstract}
A collective description of density matrix is presented for identical
multi-level atoms, which are either excited initially, driven coherently
or pumped incoherently. The density matrix is defined as expectation
value of projection or transition operators in a basis of atom's product
states. The identical matrix elements are identified with several integers,
which specify uniquely the involved operators.
To remove the redundancy, these identical elements are treated 
as single quantity and the equation for this quantity is dervied by mapping
the transition or projection operators to a single vector specified
with these integers. As a result, the number of computed elements increases polynomially
 rather than exponentially with the number of atoms. 
 As an example, we carry out exact simulation of hundreds of two-level atoms 
 and demonstrate  the different conditions for observing superradiance and superfluorescence. 

\end{abstract}
\maketitle

\section{Introduction}

The collective atom-light interaction and the resulting phenomena,
such as superradiance, have been studied intensively since the seminal
work by Dicke \citep{RHDicke} (for reviews, see \citep{AVandreev,BarryMGarraway}).
It is now established that the superradiance appears if the collective
radiative decay of atoms overcomes the decay and dephasing rate of
individual atoms \citep{AVandreev}. The superradiance is difficult
to realize with atoms in a sub-wavelength scale \citep{RHDicke} because
the dipole-dipole interaction can deteriorate correlation between
atoms. Thus, to establish long-range correlation between atoms, it
is preferable to couple atoms with single \citep{RBonifacio}
or few electromagnetic modes \citep{AVandreev-1} by, for example,
allocating them in a cylindrical shape or inside an optical cavity. 

The superradiance is often studied with laser master equation in bad-cavity
limit. In this limit, the cavity mode can be also eliminated adiabatically 
to achieve so-called superradiance master equation for the atoms,
where the coupling with the cavity mode results to the collective
decay and transition frequency (Lamb) shift of the atoms \citep{RBonifacio}.
The superradiance master equation has been solved analytically in
the basis of Dicke states \citep{CTLee} and atomic coherent states
(leading to Fokker-Planck equation) \citep{LMNarducci,RJGlauber}.
However, if we want to account for spontaneous emission or incoherent
pumping of individual atoms \citep{DMeiser,DMeiser1}, which are either
unavoidable or desired, we can only solve this equation numerically
in the basis of Dicke states \citep{BAChase,BQBaragiola,FDamanet,NShammah} or flip-spin
numbers \citep{MRichter,Ullrich,MGegg,YZhang,PKirton} {(}equivalent to SU(4)-symmetry
group theory \citep{MXu}{)}.  Here, in order to reduce the computational effort, 
 one often explores the symmetry of the master equation raising from
  the permutation symmetry of the indistinguishable atoms.  However, 
most theories focus on system with two-level atoms and it is not trivial how to generalize them to multi-level
atoms. 

\begin{figure}
\begin{centering}
\includegraphics[scale=0.75]{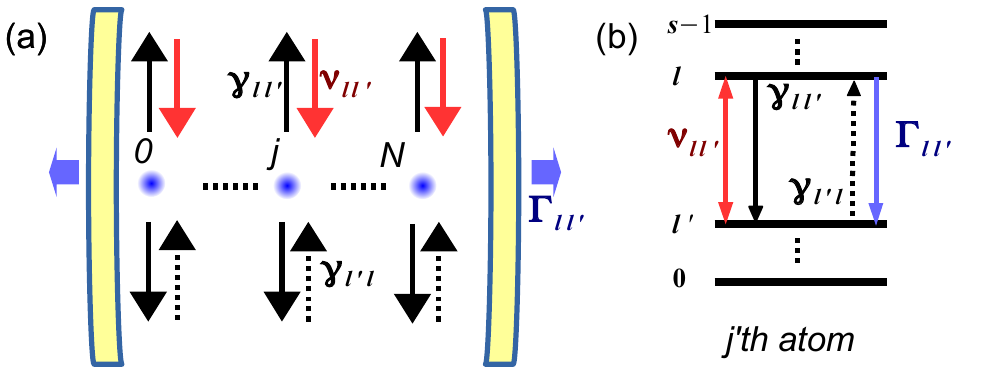}
\par\end{centering}
\caption{\label{fig:system} (a) shows identical atoms labeled by $j$ from
zero to $N$ in an optical cavity. (b) shows the atomic levels labeled
by $l$ from zero to $s-1$ and the involved processes. The atoms
might be excited coherently with a strength $v_{ll'}$ (red arrows),
pumped incoherently with a rate $\gamma_{l'l}$ (black dashed arrow),
decay individually with a rate $\gamma_{ll'}$ (black solid arrow)
or collectively with a rate $\Gamma_{ll'}$ (blue solid arrow, caused
by the coupling with the lossy cavity). For more details see text. }
\end{figure}

In this article, we present a collective description of density matrix
for identical multi-level atoms, which are either driven coherently
or pumped incoherently while subject to either individual decay or
collective decay, see Fig. \ref{fig:system} (a). In our description,
we identify the group of identical matrix elements and derive the
equation for the group, considered as single entity, by following
a rigorous procedure. Our description removes the redundancy of matrix
elements and thus is suitable for simulating superradiance from many
atoms. In essence, our description generalizes the method based on flip-spin
numbers \citep{MRichter,Ullrich,MGegg,YZhang}. More importantly,
it allows us to explore the symmetry of density matrix \citep{MXu}
without explicitly applying the group theory and thus is much
easy to be implemented. 

As an example, we utilize our description to simulate transient and
steady-state superradiance from hundreds of two-level atoms. In particular,
we correlate the radiation dynamics with the atomic dynamics. We find
that the superradiance without dipole appears from the atoms, which
are either fully excited initially or pumped incoherently. In contrast,
the superradiance with dipole (i.e. superfluorescence) appears for
the atoms, which are either prepared in superposition states or driven
coherently. In addition, we find that the superradiance is correlated
much stronger with the uncertainty of atomic angular moment. 

This article is organized as follows. In the following section, we
present our description. In Sec. \ref{sec:applications} we present
our results on the transient and steady-state superradiance from hundreds
of two-level atoms. In the end we provide some concluding remarks
and comment on the possible extensions in the future.

\section{Collective Description of Density Matrix\label{sec:collective-density-matrix}}

In this section, we present our description for identical multi-level
atoms subject to coherent driving, individual decay, pumping and collective
decay, see Fig. \ref{fig:system} (b). Such a system is described
by the master equation 
\begin{equation}
\frac{\partial }{\partial t} \rho = \frac{1}{i\hbar}\left[H_{a}+H_{d},\rho\right]-\mathcal{D}_{d}\left[\rho\right]-\mathcal{D}_{p}\left[\rho\right] + \frac{1}{i\hbar}\left[H_{s},\rho\right] -\mathcal{D}_{c}\left[\rho\right] \label{eq:meq}
\end{equation}
for the density operator $\rho$.  The first three terms on the right side of Eq. (\ref{eq:meq}) describe the processes related 
to individual atoms. $H_{a}=\sum_{l=0}^{s-1}\hbar\omega_{l} \sum_{j=1}^N \left|l_{j}\right\rangle \left\langle l_{j}\right|$ describes the Hamiltonian of 
$N$ identical atoms, which are labeled by $j$ and have $s$ states (levels) $\left|l_{j}\right\rangle$ with energies  $\hbar\omega_{l}$ each, see Fig. \ref{fig:system}(b). To simplify the notion, we introduce the operator $\sigma^j_{ll'}=\left|l_{j}\right\rangle \left\langle l'_{j}\right|$
for individual atoms and then the collective operator  $\sigma_{ll'}=\sum_{j}  \sigma^j_{ll'}$ for the atom ensemble.  
As a result, we can rewrite $H_{a}$ as $\sum_l\hbar\omega_l \sigma_{ll}$.  
$H_{d}=\hbar\sum_{l>l'}\left(v_{ll'}\sigma_{ll'}+\mathrm{h.c.}\right)$
describes the interaction between the $l-l'$ transitions of the atoms
and an external field of frequency $\omega_{d}$ with the strength
$v_{ll'}=v_{ll'}^{0}\exp\left(-i\omega_{d}t\right)$ (in rotating-wave
approximation). $\mathcal{D}_{d}\left[\rho\right]=\sum_{l\neq l'}\frac{\gamma_{ll'}}{2}\sum_{j}\mathcal{D}\left[\sigma_{l'l}^{j}\right]\rho$
describes the dissipation of individual atoms, which includes either
the decay from the upper level ($l>l'$) or the incoherent pumping
\citep{DMeiser,DMeiser1} from the lower level ($l<l'$). $\mathcal{D}_{p}\left[\rho\right]=\sum_{l>l'}\frac{\xi_{ll'}}{2}\sum_{j}\mathcal{D}\left[\sigma_{ll}^{j}-\sigma_{l'l'}^{j}\right]\rho$
describes the dephasing $\xi_{ll'}$ between $l-l'$ transition \footnote{The dephasing of two-level atoms is normally described by the Lindblad term
$\frac{\xi_{10}}{2}\sum_{j}\mathcal{D}\left[\sigma_{z}\right]\rho$
with the Pauli operator $\sigma_{z}=\sigma_{11}^{j}-\sigma_{00}^{j}$. Similarly, we can view any pair of levels as a two-level system and introduce
the dephasing for these levels by replacing $\sigma_{z}^{j}$ with
$\sigma_{ll}^{j}-\sigma_{l'l'}^{j}$. }. Here, the superoperator is defined as $\mathcal{D}\left[o\right]=\left\{ o^{+}o,\rho\right\} -2o\rho o^{+}$
(for any operator $o$). 

The remaining terms on the right side of Eq. (\ref{eq:meq}) are obtained by adiabatically
eliminating the cavity mode from the laser master equation. The Hamiltonian
$H_{s}=\sum_{l>l'}\hbar\Omega_{ll'}\sigma_{ll'}\sigma_{l'l}$ describes
the collective Lamb shift of the atomic transitions $\Omega_{ll'}=\left|g_{ll'}\right|^{2}\chi_{ll'}\left[\chi_{ll'}^{2}+\left(\kappa/2\right)^{2}\right]^{-1}$and
the Lindblad term $\mathcal{D}_{c}\left[\rho\right]=\sum_{l>l'}\frac{\Gamma_{ll'}}{2}\mathcal{D}\left[\sigma_{l'l}\right]\rho$
describes the collective atomic decay $\Gamma_{ll'}=\left|g_{ll'}\right|^{2}\left(\kappa/2\right)\left[\chi_{ll'}^{2}+\left(\kappa/2\right)^{2}\right]^{-1}$.
Here, $\omega_{c},\kappa$ are the cavity mode frequency and loss
rate, respectively, $\chi_{ll'}=\omega_{l}-\omega_{l'}-\omega_{c},g_{ll'}$
are the frequency detuning and the coupling between the atoms and the
cavity mode.

To solve the master equation (\ref{eq:meq}) we introduce the product states $\left|\alpha\right\rangle \equiv\prod_{j=1}^{N}\left|l_{j}\right\rangle $
and $\left|\beta\right\rangle \equiv\prod_{j=1}^{N}\left|l'_{j}\right\rangle $
with the sets $\alpha=\left\{ l_{1},...,l_{N}\right\} $, $\beta=\left\{ l'_{1},...,l'_{N}\right\} $
and then define the density matrix elements $\rho_{\beta\alpha}$
as the expectation value $\mathrm{tr}\left\{ \rho\left|\alpha\right\rangle \left\langle \beta\right|\right\} $
of the transition ($\alpha\neq\beta$) or projection ($\alpha=\beta$)
operator $\left|\alpha\right\rangle \left\langle \beta\right|$. If
all the atoms are identical we expect that many elements are identical
and thus do not need to consider all of them. In order to do so,
we need a clever way to identify these elements. To this end we should analyze
the equation for the density matrix elements $\frac{\partial}{\partial t}\rho_{\beta\alpha}=\mathrm{tr}\left\{ \left(\frac{\partial}{\partial t}\rho\right)\left|\alpha\right\rangle \left\langle \beta\right|\right\} $.
If we utilize the master equation \cob{ (\ref{eq:meq})} for $\rho$ and then cyclize the operators
in the expectation values such that $\rho$ appears always on the
right side, we can then rewrite the equation as $\frac{\partial}{\partial t}\rho_{\beta\alpha}=\mathrm{tr}\left\{ \left(\frac{\partial}{\partial t}\left|\alpha\right\rangle \left\langle \beta\right|\right)\rho\right\} $
with an ancillary equation $\frac{\partial}{\partial t}\left|\alpha\right\rangle \left\langle \beta\right|\equiv\left(i/\hbar\right)\left[H_{a}+H_{d}+H_{s},\left|\alpha\right\rangle \left\langle \beta\right|\right]-\tilde{\mathcal{D}}_{d}\left[\left|\alpha\right\rangle \left\langle \beta\right|\right]-\tilde{\mathcal{D}}_{p}\left[\left|\alpha\right\rangle \left\langle \beta\right|\right]-\tilde{\mathcal{D}}_{c}\left[\left|\alpha\right\rangle \left\langle \beta\right|\right]$,
where the latter three terms are defined with the superoperator $\tilde{\mathcal{D}}\left[o\right]=\left\{ o^{+}o,\left|\alpha\right\rangle \left\langle \beta\right|\right\} -2o^{+}\left|\alpha\right\rangle \left\langle \beta\right|o$.
Here, we switch $o^{+}$ and $o$ in the last term compared
to $\mathcal{D}\left[o\right]$. We should not consider this equation
as an equation in Heisenberg picture but should consider it  only as a tool
to analyze the equation for $\rho_{\beta\alpha}$. 

In the ancillary equation, we encounter two kinds of terms. One kind
has the form like $\sum_{j}\sigma_{ll'}^{j}\left|\alpha\right\rangle \left\langle \beta\right|,\left|\alpha\right\rangle \left\langle \beta\right|\sum_{j}\sigma_{ll'}^{j}$
while another kind has the form like $\sum_{j}\sigma_{ll'}^{j}\left|\alpha\right\rangle \left\langle \beta\right|\sigma_{l'l}^{j}$.
Using the identity operator $\sum_{k}\sigma_{kk}^{j}$ we can always
write the first kind as a sum of the second kind. In general, we should
evaluate the sandwich structure $\sum_{j}\sigma_{ll'}^{j}\left|\alpha\right\rangle \left\langle \beta\right|\sigma_{kk'}^{j}$.
We can do so in two steps. First, we evaluate whether the $j$'th
atom is on the state $\left|l'_{j}\right\rangle $ (from $\sigma_{ll'}^{j}$)
in the product state $\left|\alpha\right\rangle $ and simultaneously
on the state $\left|k_{j}\right\rangle $ (from $\sigma_{kk'}^{j}$)
in the product state $\left|\beta\right\rangle $. Second, if the
evaluation turns out to be true, we replace $\left|l'_{j}\right\rangle $
by $\left|l_{j}\right\rangle $ in $\left|\alpha\right\rangle $ and
$\left|k_{j}\right\rangle $ by $\left|k'_{j}\right\rangle $ in $\left|\beta\right\rangle $
to obtain two new product states $\left|\alpha'\right\rangle $ and
$\left|\beta'\right\rangle $ , which form a new transition
or projection operator $\left|\alpha'\right\rangle \left\langle \beta'\right|$.
If we repeat the two steps for all the atoms, we  obtain $n_{l'k}$
terms, where $n_{l'k}$ are the number of atoms fulfilling the evaluation
in the first step. If all the atoms are identical, we expect the following relation for the 
expectation values 
\begin{equation}
{\rm tr} \{\rho \sum_{j}\sigma_{ll'}^{j}\left|\alpha\right\rangle \left\langle \beta\right|\sigma_{kk'}^{j} \}={\rm tr} \{\rho n_{ll'}  \left|\alpha'\right\rangle \left\langle \beta'\right| \}.
\end{equation}
To facilitate the following derviation, we can ignore the trace and the 
density operator temporarily and assume a relation for the operators
\begin{equation}
 \sum_{j}\sigma_{ll'}^{j}\left|\alpha\right\rangle \left\langle \beta\right|\sigma_{kk'}^{j} = n_{ll'}  \left|\alpha'\right\rangle \left\langle \beta'\right| .\label{eq:sandwich-structure}
\end{equation}

\begin{figure}
\begin{centering}
\includegraphics[scale=0.37]{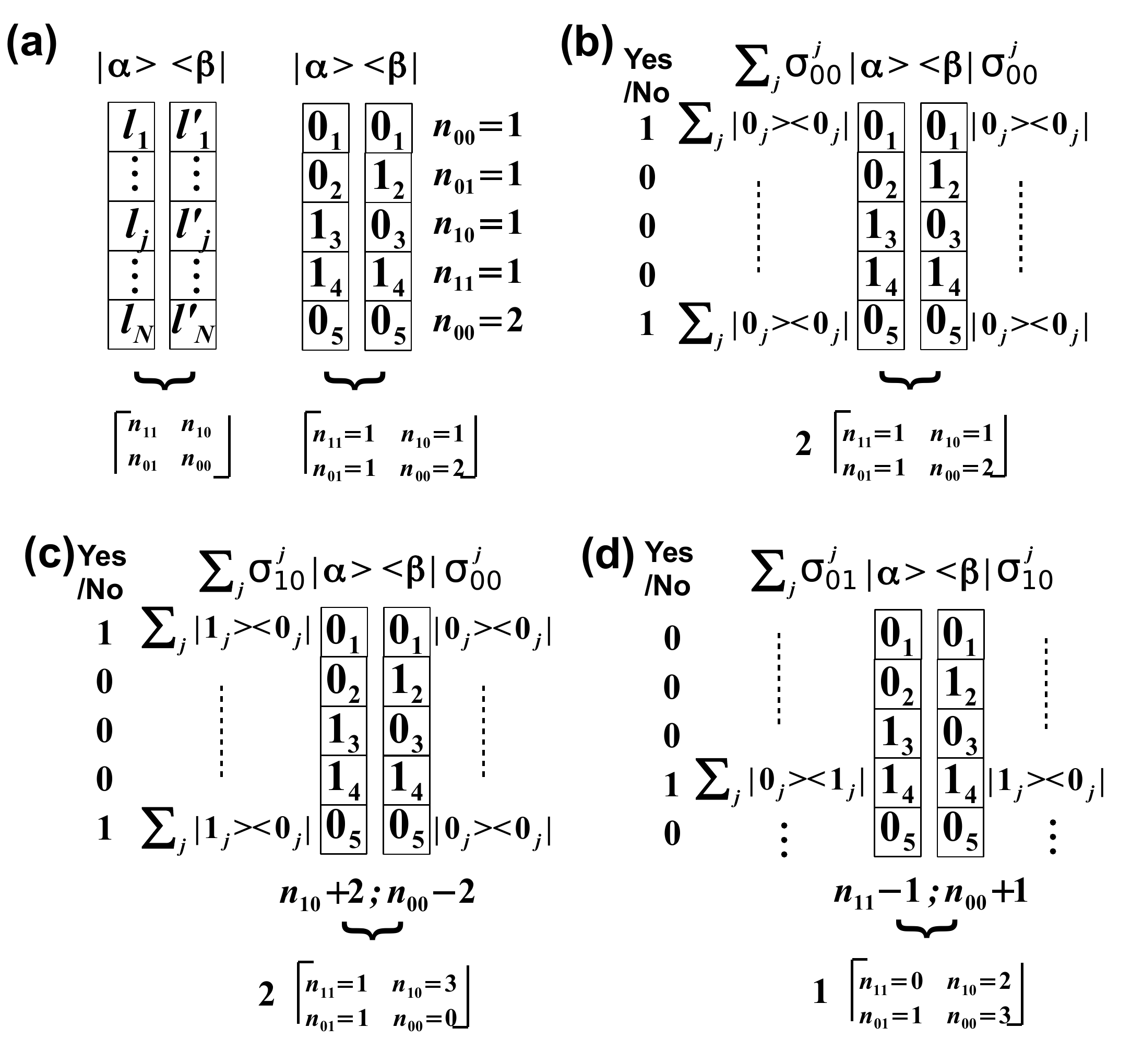}
\par\end{centering}
\caption{\label{fig:actions} Mapping of projection or transition operators
$\left|\alpha\right\rangle \left\langle \beta\right|$ to single operators
$\left\lceil n\right\rfloor $ (a) and three kinds of actions of the
superoperators on these operators $\left\lceil n\right\rfloor $ (b,c,d)
for the exemplary system with five identical two-level atoms. For
more details, see the text. }
\end{figure}

If we apply Eq. (\ref{eq:sandwich-structure}) to those operators
$\left|\alpha\right\rangle \left\langle \beta\right|$, which lead
to the elements $\rho_{\beta\alpha}$ with same value, we encounter
several integers $n_{ll'}$ ($l,l'=0,...s-1$). In other words, the
set of integers $\left\{ n_{ll'}\right\} $ specify uniquely the group
(or set) of operators $\left\{ \left|\alpha\right\rangle \left\langle \beta\right|\right\} $,
where the members $\left|\alpha\right\rangle \left\langle \beta\right|$
are affected in the same way by the sandwich structures (and thus
the equation $\frac{\partial}{\partial t}\left|\alpha\right\rangle \left\langle \beta\right|$).
Using this finding we can remove the redundancy of matrix elements.
In the following, we present a rigorous procedure for doing so.

The core of our procedure is to define an operator $\left\lceil n\right\rfloor $
with the set of integers $n\equiv\left\{ n_{ll'}\right\} $ to represent
any member of the group $\left\{ \left|\alpha\right\rangle \left\langle \beta\right|\right\} $.
In other words, we establish a many-to-one mapping 
\begin{equation}
\left|\alpha\right\rangle \left\langle \beta\right|\leftrightarrow\left\lceil n\right\rfloor =\left\lceil \begin{array}{ccc}
n_{s-1s-1} & \cdots & n_{s0}\\
\vdots & \ddots & \vdots\\
n_{0s} & \cdots & n_{00}
\end{array}\right\rfloor .\label{eq:mapping-multi-levels}
\end{equation}
Fig. \ref{fig:actions}(a) shows one example of the mapping for two-level atoms.
Here, we choose the matrix-like form $\left\lceil \right\rfloor $
to reflect $\left\lceil n\right\rfloor $ as operators in \emph{Hilbert}
space. Furthermore, we can define the expectation value $\left\langle n\right\rangle \equiv\mathrm{tr}\left\{ \rho\left\lceil n\right\rfloor \right\} $
and then establish a similar mapping between the matrix elements $\rho_{\beta\alpha}$
and $\left\langle n\right\rangle $. If we replace $\left|\alpha\right\rangle \left\langle \beta\right|$
with $\left\lceil n\right\rfloor $ in the ancillary equation $\frac{\partial}{\partial t}\left|\alpha\right\rangle \left\langle \beta\right|$
and then take a trace after multiplying with $\rho$, i.e. $\left\langle \cdot\right\rangle =\mathrm{tr}\left\{\rho \cdot\right\} $,
on both sides, we get the following equation for $\left\langle n\right\rangle $:
\begin{align}
 & \frac{\partial}{\partial t}\left\langle n\right\rangle =\left(i/\hbar\right)\left\langle \left[H_{a}+H_{d}+H_{s},\left\lceil n\right\rfloor \right]\right\rangle \nonumber \\
 & -\sum_{l\neq l'}\frac{\gamma_{ll'}}{2}\Bigl\langle\left\{ \sigma_{ll},\left\lceil n\right\rfloor \right\} -2\sum_{j}\sigma_{ll'}^{j}\left\lceil n\right\rfloor \sigma_{l'l}^{j}\Bigr\rangle\nonumber \\
 & -\sum_{l>l'}\frac{\xi_{ll'}}{2}\Bigl\langle\left\{ \sigma_{ll}-\sigma_{l'l'},\left\lceil n\right\rfloor \right\} -2\sum_{j}\left(\sigma_{ll}^{j}-\sigma_{l'l'}^{j}\right)\left\lceil n\right\rfloor \left(\sigma_{ll}^{j}-\sigma_{l'l'}^{j}\right)\Bigr\rangle\nonumber \\
 & -\sum_{l>l'}\frac{\Gamma_{ll'}}{2}\Bigl\langle\Bigl\{\sigma_{ll'}\sigma_{l'l},\left\lceil n\right\rfloor \Bigr\}-2\sigma_{ll'}\left\lceil n\right\rfloor \sigma_{l'l}\Bigr\rangle.\label{eq:cdm-multi-levels}
\end{align}
In the second line, we have utilized $\sigma_{ll'}^{j}\sigma_{l'l}^{j}=\sigma_{ll}^{j}$
and $\sigma_{ll}=\sum_{j}\sigma_{ll}^{j}$. Since $\left\langle n\right\rangle $
represents the group of $\rho_{\beta\alpha}$ of same value, we might
call it as \emph{collective density matrix}. In the following, we
evaluate the different terms in the above equation. 

\subsection{Contributions to Collective Density Matrix Equation\label{subsec:collective-master-equation}}

To evaluate the contribution $\left(\frac{\partial}{\partial t}\left\langle n\right\rangle \right)_{a}$
of $H_{a}$ in Eq. (\ref{eq:cdm-multi-levels}), we encounter two
terms $\sigma_{ll}\left\lceil n\right\rfloor $,$\left\lceil n\right\rfloor \sigma_{ll}$.
To evaluate them we consider $\left\lceil n\right\rfloor $ as a vector
in Liouville space and promote $\sigma_{ll}$ as a superoperator,
which can act on $\left\lceil n\right\rfloor $ either from the left side $\sigma_{ll}\left\lceil n\right\rfloor $
or from the right side $\left\lceil n\right\rfloor \sigma_{ll}$. As an example, we demonstrate 
the evaluation of $\sigma_{ll}\left\lceil n\right\rfloor $ . Using the identity operator $\sum_{k}\sigma_{kk}^{j}$
of the $j$'th atom, we can rewrite $\sigma_{ll}\left\lceil n\right\rfloor $ as $\sum_{k}\sum_{j}\sigma_{ll}^{j}\left\lceil n\right\rfloor \sigma_{kk}^{j}.$
Here, we have expanded $\sigma_{ll}$ as $\sum_{j}\sigma_{ll}^{j}$.
To proceed, we turn $\left\lceil n\right\rfloor $ back to $\left|\alpha\right\rangle \left\langle \beta\right|$
and then apply Eq. (\ref{eq:sandwich-structure}) to obtain $\sum_{j}\sigma_{ll}^{j}\left|\alpha\right\rangle \left\langle \beta\right|\sigma_{kk}^{j}=n_{lk}\left|\alpha\right\rangle \left\langle \beta\right|$.
Notice that the product states do not change on the right side. Using
the mapping of Eq. (\ref{eq:mapping-multi-levels}), we obtain 
\begin{equation}
\sum_{j}\sigma_{ll}^{j}\left\lceil n\right\rfloor \sigma_{kk}^{j}=n_{lk}\left\lceil n\right\rfloor .\label{eq:relation}
\end{equation}
This relation is exemplified with Fig. \ref{fig:actions} (b) for
two-level atoms. If we sum up the above relations for different $k$, we obtain
\begin{equation}
\sigma_{ll}\left\lceil n\right\rfloor =\sum_{l'}n_{ll'}\left\lceil n\right\rfloor .\label{eq:ll-left}
\end{equation}
We can follow the same procedure to evaluate $\left\lceil n\right\rfloor \sigma_{ll}$.
Notice that this time we should apply the identity operator from the
left side. As a result, we should obtain 
\begin{equation}
\left\lceil n\right\rfloor \sigma_{ll}=\sum_{l'}n_{l'l}\left\lceil n\right\rfloor .\label{eq:ll-right}
\end{equation}
Using the above expressions, we finally get the contribution of $H_{a}$
to the master equation
\begin{equation}
\left(\frac{\partial}{\partial t}\left\langle n\right\rangle \right)_{a}=i\sum_{l}\omega_{l}\sum_{l'}\left(n_{ll'}-n_{l'l}\right)\left\langle n\right\rangle .\label{eq:three-level-atoms-0}
\end{equation}

To evaluate the contribution $\left(\frac{\partial}{\partial t}\left\langle n\right\rangle \right)_{d}$
of $H_{d}$ in Eq. (\ref{eq:cdm-multi-levels}), we have to evaluate
the action of $\sigma_{ll'}$ ($l\neq l'$) on $\left\lceil n\right\rfloor $
from the left side $\sigma_{ll'}\left\lceil n\right\rfloor $ and
from the right side $\left\lceil n\right\rfloor \sigma_{ll'}$. As an example,  we 
detail the evalulation of $\sigma_{ll'}\left\lceil n\right\rfloor $. Using the identity operator of the $j$-th atom,
we can rewrite this term as $\sum_{k}\sum_{j}\sigma_{ll'}^{j}\left\lceil n\right\rfloor \sigma_{kk}^{j}$.
To evaluate the terms inside the sum to $k$, we can turn $\left\lceil n\right\rfloor $
back to $\left|\alpha\right\rangle \left\langle \beta\right|$ and
then apply Eq. (\ref{eq:sandwich-structure}) to obtain $\sum_{j}\sigma_{ll'}^{j}\left|\alpha\right\rangle \left\langle \beta\right|\sigma_{kk}^{j}=n_{l'k}\left|\alpha'\right\rangle \left\langle \beta\right|$.
Notice that the product state $\left|\beta\right\rangle $ does not
change on the right side. The product state $\left|\alpha'\right\rangle $
differs from $\left|\alpha\right\rangle $ by that one of the atoms,
which is initially on $\left|l'\right\rangle $ state, is now on $\left|l\right\rangle $
state. Using the mapping of Eq. (\ref{eq:mapping-multi-levels}),
we can map $\left|\alpha'\right\rangle \left\langle \beta\right|$
to a new operator $\left\lceil n'\right\rfloor $, which differs from
$\left\lceil n\right\rfloor $ by that $n_{l'k}$ reduces by one and
$n_{lk}$ increases by one. To sum up, we establish the following
relation 
\begin{align}
\sigma_{ll'}\left\lceil n\right\rfloor  & =\sum_{k}n_{l'k}\left\lceil \left\{ n_{l'k}-1,n_{lk}+1\right\} \right\rfloor .\label{eq:ll'-left}
\end{align}
Here and in the following, for simplicity, we indicate the new operator
$\left\lceil n'\right\rfloor $ with only the changed numbers relative
to $\left\lceil n\right\rfloor $. Fig. \ref{fig:actions}(c) shows
one example of the above relation for two-level atoms. Following
the same procedure, we can also get 
\begin{align}
\left\lceil n\right\rfloor \sigma_{ll'} & =\sum_{k}n_{kl}\left\lceil \left\{ n_{kl}-1,n_{kl'}+1\right\} \right\rfloor .\label{eq:ll'-right}
\end{align}
Using the above expressions we get the contribution of $H_{d}$ to
the master equation 

\begin{align}
 & \left(\frac{\partial}{\partial t}\left\langle n\right\rangle \right)_{d}=\nonumber \\
 & i\sum_{l\neq l'}v_{ll'}\sum_{k}\left(n_{l'k}\left\langle n_{l'k}-1,n_{lk}+1\right\rangle -n_{kl}\left\langle n_{kl}-1,n_{kl'}+1\right\rangle \right)\nonumber \\
 & +i\sum_{l\neq l'}v_{ll'}^{*}\sum_{k}\left(n_{lk}\left\langle n_{lk}-1,n_{l'k}+1\right\rangle -n_{kl'}\left\langle n_{kl'}-1,n_{kl}+1\right\rangle \right).\label{eq:three-level-interaction}
\end{align}

To evaluate the contribution $\left(\frac{\partial}{\partial t}\left\langle n\right\rangle \right)_{s}$
of $H_{s}$ in Eq. (\ref{eq:cdm-multi-levels}), we have to evaluate
the action of $\sigma_{ll'}\sigma_{l'l}$ ($l\neq l'$) on $\left\lceil n\right\rfloor $
from the left side $\sigma_{ll'}\sigma_{l'l}\left\lceil n\right\rfloor $
and from the right side $\left\lceil n\right\rfloor \sigma_{ll'}\sigma_{l'l}$.
To evaluate these terms we apply simply Eq. (\ref{eq:ll'-left}) and
(\ref{eq:ll'-right}) twice to obtain 
\begin{align}
 & \sigma_{ll'}\sigma_{l'l}\left\lceil n\right\rfloor =\sum_{k}n_{lk}\left(n_{l'k}+1\right)\left\lceil n\right\rfloor \nonumber \\
 & +\sum_{k\neq k'}n_{lk}n_{l'k'}\left\lceil \left\{ n_{lk}-1,n_{l'k'}-1,n_{l'k}+1,n_{l'k'}+1\right\} \right\rfloor ,\label{eq:ll'-left-left}\\
 & \left\lceil n\right\rfloor \sigma_{ll'}\sigma_{l'l}=\sum_{k}n_{kl}\left(n_{kl'}+1\right)\left\lceil n\right\rfloor \nonumber \\
 & +\sum_{k\neq k'}n_{kl}n_{k'l'}\left\lceil \left\{ n_{kl}-1,n_{k'l'}-1,n_{kl'}+1,n_{k'l}+1\right\} \right\rfloor .\label{eq:ll'-right-right}
\end{align}
Notice that the operators in the second and fourth line differ from
$\left\lceil n\right\rfloor $ by four numbers as indicated. Using
these expressions we obtain 
\begin{align}
 & \left(\frac{\partial}{\partial t}\left\langle n\right\rangle \right)_{s}=i\sum_{l>l'}\Omega_{ll'}\Bigl[ \sum_{k}\left(n_{lk}\left(n_{l'k}+1\right)-n_{kl}\left(n_{kl'}+1\right)\right)\left\langle n\right\rangle \nonumber \\
 & +\sum_{k\neq k'}n_{lk}n_{l'k'}\left\langle n_{lk}-1,n_{l'k'}-1,n_{l'k}+1,n_{l'k'}+1\right\rangle \nonumber \\
 & -\sum_{k\neq k'}n_{kl}n_{k'l'}\left\langle n_{kl}-1,n_{k'l'}-1,n_{kl'}+1,n_{k'l}+1\right\rangle \Bigr].
\end{align}

To evaluate the contribution $\left(\frac{\partial}{\partial t}\left\langle n\right\rangle \right)_{d}$
of the individual dissipation $\mathcal{D}_{d}\left[\rho\right]$
in Eq. (\ref{eq:cdm-multi-levels}), we encounter $\sigma_{ll}\left\lceil n\right\rfloor ,\left\lceil n\right\rfloor \sigma_{ll}$
and $\sum_{j}\sigma_{ll'}^{j}\left\lceil n\right\rfloor \sigma_{l'l}^{j}$
. Since the former two expressions can be evaluated with Eq. (\ref{eq:ll-left})
and (\ref{eq:ll-right}), we  focus on the last term. In the same
spirit as before, we turn $\left\lceil n\right\rfloor $ back to $\left|\alpha\right\rangle \left\langle \beta\right|$
and then apply Eq. (\ref{eq:sandwich-structure}) to obtain $\sum_{j}\sigma_{ll'}^{j}\left|\alpha\right\rangle \left\langle \beta\right|\sigma_{l'l}^{j}=n_{l'l'}\left|\alpha'\right\rangle \left\langle \beta'\right|$.
Here, the new product state $\left|\alpha'\right\rangle $ ($\left|\beta'\right\rangle $)
differs from $\left|\alpha\right\rangle $ ($\left|\beta\right\rangle $)
by that one of the atoms, which is on $\left|l'\right\rangle $ state
in $\left|\alpha\right\rangle $ ($\left|\beta\right\rangle $), is
now on $\left|l\right\rangle $ state. Using the mapping of Eq. (\ref{eq:mapping-multi-levels}),
we map $\left|\alpha'\right\rangle \left\langle \beta'\right|$ to
a new operator $\left\lceil n'\right\rfloor $, which differs from
$\left\lceil n\right\rfloor $ by that $n_{l'l'}$ reduces by one
and $n_{ll}$ increases by one. To sum up, we establish the following
relation 
\begin{equation}
\sum_{j}\sigma_{ll'}^{j}\left\lceil n\right\rfloor \sigma_{l'l}^{j}=n_{l'l'}\left\lceil \left\{ n_{l'l'}-1,n_{ll}+1\right\} \right\rfloor .\label{eq:ll-l'l'}
\end{equation}
This relation is exemplified with Fig. \ref{fig:actions}(d) for two-level
atoms. With the above expression and also Eqs. (\ref{eq:ll-left})
and (\ref{eq:ll-right}) we get 
\begin{equation}
\left(\frac{\partial}{\partial t}\left\langle n\right\rangle \right)_{d}=\sum_{l\neq l'}\frac{\gamma_{ll'}}{2}\Bigl[\sum_{k}\left(n_{lk}+n_{kl}\right)\left\langle n\right\rangle -2n_{l'l'}\left\langle n_{l'l'}-1,n_{ll}+1\right\rangle \Bigr].\label{eq:individual-disspition}
\end{equation}

To evaluate the contribution $\left(\frac{\partial}{\partial t}\left\langle n\right\rangle \right)_{p}$
of the individual dephasing $\mathcal{D}_{p}\left[\rho\right]$ in
Eq. (\ref{eq:cdm-multi-levels}), we encounter the terms $\sigma_{ll}\left\lceil n\right\rfloor, \left\lceil n\right\rfloor \sigma_{ll}$,$\sum_{j}\sigma_{ll}^{j}\left\lceil n\right\rfloor \sigma_{ll}^{j}, \sum_{j}\sigma_{ll}^{j}\left\lceil n\right\rfloor \sigma_{l'l'}^{j}$
(also terms with $l,l'$ exchanged). The former two terms can be evaluated
with Eq. (\ref{eq:ll-left}) and (\ref{eq:ll-right}). The latter
two terms can be evaluated with Eq. (\ref{eq:relation}). As a result,
we get 
\begin{align}
\left(\frac{\partial}{\partial t}\left\langle n\right\rangle \right)_{p} & =-\sum_{l>l'}\frac{\xi_{ll'}}{2}\Bigl[\sum_{k}\left(n_{lk}+n_{kl}-n_{l'k}-n_{kl'}\right)\nonumber \\
 & -2\left(n_{ll}-n_{ll'}-n_{l'l}+n_{l'l'}\right)\Bigr]\left\langle n\right\rangle .
\end{align}

To evaluate the contribution $\left(\frac{\partial}{\partial t}\left\langle n\right\rangle \right)_{c}$
of the collective decay $\mathcal{D}_{c}\left[\rho\right]$ in Eq.
(\ref{eq:cdm-multi-levels}) we encounter three terms $\sigma_{ll'}\sigma_{l'l}\left\lceil n\right\rfloor $,
$\left\lceil n\right\rfloor \sigma_{ll'}\sigma_{l'l}$ and $\sigma_{l'l}\left\lceil n\right\rfloor \sigma_{ll'}$.
To evaluate the former two terms, we can simply apply Eq. (\ref{eq:ll'-left})
and (\ref{eq:ll'-right}) twice to get Eq. (\ref{eq:ll'-left-left})
and \ref{eq:ll'-right-right}. In order to evaluate the last term,
we apply Eq. (\ref{eq:ll'-left}) and (\ref{eq:ll'-right}) once to
get
\begin{align}
 & \sigma_{ll'}\left\lceil n\right\rfloor \sigma_{l'l}=n_{l'l'}\left(n_{l'l}+1\right)\left\lceil \left\{ n_{l'l'}-1,n_{ll}+1\right\} \right\rfloor \nonumber \\
 & +\sum_{k\neq l',k'\neq l}n_{kl'}n_{l'k'}\left\lceil \left\{ n_{kl'}-1,n_{l'k'}-1,n_{kl}+1,n_{lk'}+1\right\} \right\rfloor .
\end{align}
With the above expressions, we finally obtain 
\begin{align}
 & \left(\frac{\partial}{\partial t}\left\langle n\right\rangle \right)_{c}=-\sum_{l>l'}\frac{\Gamma_{ll'}}{2}\Bigl\{\sum_{k}\left[n_{lk}\left(n_{l'k}+1\right)+n_{kl}\left(n_{kl'}+1\right)\right]\left\langle n\right\rangle \nonumber \\
 & +\sum_{k\neq k'}\Bigl(n_{lk}n_{l'k'}\left\langle n_{lk}-1,n_{l'k'}-1,n_{l'k}+1,n_{l'k'}+1\right\rangle \nonumber \\
 & +n_{kl}n_{k'l'}\left\langle n_{kl}-1,n_{k'l'}-1,n_{kl'}+1,n_{k'l}+1\right\rangle \Bigr)\nonumber \\
 & -2\Bigl[n_{l'l'}\left(n_{l'l}+1\right)\left\langle n_{l'l'}-1,n_{ll}+1\right\rangle \nonumber \\
 & +\sum_{k\neq l',k'\neq l}n_{kl'}n_{l'k'}\left\langle n_{kl'}-1,n_{l'k'}-1,n_{kl}+1,n_{lk'}+1\right\rangle \Bigr]\Bigr\}.
\end{align}

\subsection{Initial Conditions and Observables\label{subsec:initial-conditions}}

In the above subsection, we have obtained self-consistent equation
for the collective density matrix $\left\langle n\right\rangle $.
To solve this equation, we have to also specify the initial condition.
In general, the initial states of individual atoms can be mixed states
and thus should be specified by a set of pure states $\left\{ \left|\psi_{i}\right\rangle _{j}\right\} $
and a set of probabilities $\left\{ m_{i}\right\} $ (satisfying $\sum_{i}m_{i}=1$).
From the two sets we obtain the initial density operator $\rho_{0,j}\equiv\sum_{i}m_{i}\left|\psi_{i}\right\rangle _{j}\left\langle \psi_{i}\right|_{j}$
of the $j$'th atom. The pure states $\left|\psi_{i}\right\rangle _{j}$
can be specified as $\left|\psi_{i}\right\rangle _{j}=\sum_{l}c_{l}^{i}\left|l_{j}\right\rangle $
with complex numbers $c_{j}^{i}$ (fulfilling the condition $\sum_{l}\left|c_{l}^{i}\right|^{2}=1$).
Here, we assume that $m_{i}$ and $c_{l}^i$ do not depend on the specific
atoms. If there is no any correlation between the atoms, we can obtain
the initial density operator of all the atoms with the direct product
$\rho_{0}=\prod_{j}\rho_{0,j}$.

To proceed, we insert the definition of $\rho_{0,j}$ to $\rho_{0}$
and then combine the states $\bigl|l_{j}\bigr\rangle$ ($\bigl\langle l'_{j}\bigr|$)
on the left (right) side to form the product states $\left|\beta\right\rangle $
($\left\langle \alpha\right|$), and then map the operator $\left|\beta\right\rangle \left\langle \alpha\right|$
to the operator $\left\lceil n\right\rfloor $. As a result, we get
the initial density operator of the atoms 
\begin{equation}
\rho_{0}=\sum_{\left\{ n_{ll'}\right\} }C_{\left\{ n_{ll'}\right\} }\prod_{l,l'}\left(\sum_{i}m_{i}c_{l}^{i}c_{l'}^{i*}\right)^{n_{ll'}}\left\lceil n\right\rfloor .\label{eq:initial-density-operator}
\end{equation}
Here, $\sum_{\left\{ n_{ll'}\right\} }$ indicates the sum to those
$n_{ll'}$ satisfying the relation $\sum_{ll'}n_{ll'}=N$. $C_{\left\{ n_{ll'}\right\} }$
is the number \footnote{The number $C_{\left\{ n_{ll'}\right\} }$ can be computed as follows.
We imagine the atoms as balls and $\left\lceil n\right\rfloor $ as
a collection of $s^{2}$ boxes labeled by $ll'$. Then, this number is equivalent
to the number of possibilities that $N$ balls are arranged in the $s^{2}$
boxes such that $n_{ll'}$ balls fall in the box $ll'$. After some
calculation we get the number $C_{\left\{ n_{ll'}\right\} }=C_{N}^{n_{00}}C_{N-n_{00}}^{n_{10}}...C_{N-\sum_{ll'}\left(1-\delta_{ll',s-1s-1}\right)n_{ll'}}^{n_{s-1s-1}}$
with the binomial function $C_{n}^{m}=n!/\left[m!\left(n-m\right)!\right]$. } of $\left|\beta\right\rangle \left\langle \alpha\right|$ mapped
to the same $\left\lceil n\right\rfloor $. With the above expression,
we can obtain the initial condition for the collective density matrix
$\left\langle n\right\rangle _{0}=\mathrm{tr}\left\{ \rho_{0}\left\lceil n\right\rfloor \right\} $:
\begin{equation}
\left\langle n\right\rangle _{0}=\prod_{l,l'=0}^{s-1}\left(\sum_{i}m_{i}c_{l}^{i}c_{l'}^{i*}\right)^{n_{ll'}}.\label{eq:initial-values}
\end{equation}

Once we solve the equation for the collective density matrix, we would
like to analyze the system dynamics. To this end, we have to compute
the observable of interest. In general, the observable is expectation
value $\left\langle O\right\rangle =\mathrm{tr}\left\{ O\rho\right\} $
of a collective operator $O=\sum_{j}o_{j}$. Using the identity operator
$\sum_{\alpha}\left|\alpha\right\rangle \left\langle \alpha\right|$,
we can rewrite the observable as $\left\langle O\right\rangle =\mathrm{tr}\left\{ O\sum_{\alpha}\left|\alpha\right\rangle \left\langle \alpha\right|\rho\right\} $.
To proceed, we consider the mapping of the projection operator 
\begin{equation}
\left|\alpha\right\rangle \left\langle \alpha\right|\leftrightarrow\left\lceil \left\{ n_{ll}\right\} \right\rfloor \equiv\left\lceil \begin{array}{ccc}
n_{s-1s-1} & 0 & 0\\
0 & \ddots & 0\\
0 & 0 & n_{00}
\end{array}\right\rfloor .
\end{equation}
Here, $n_{ll}$ denotes the number of atoms on the $\left|l\right\rangle $
states in the product states $\left|\alpha\right\rangle $. Notice
that the numbers $n_{ll'}$ with $l\neq l'$ are zero. To simplify
the notation we utilize $\left\lceil \left\{ n_{ll}\right\} \right\rfloor $
to represent this specific operator. Using the same argument as applied
to obtain Eq. (\ref{eq:initial-density-operator}), we can obtain the relation
\begin{equation}
\sum_{\alpha}\left|\alpha\right\rangle \left\langle \alpha\right|=\sum_{\left\{ n_{ll}\right\} }C_{\left\{ n_{ll}\right\} }\left\lceil \left\{ n_{ll}\right\} \right\rfloor ,\label{eq:identity}
\end{equation}
where $\sum_{\left\{ n_{ll}\right\} }$ is the sum to $n_{ll}$ satisfying
$\sum_{l}n_{ll}=N$ and $C_{\left\{ n_{ll}\right\} }=\prod_{k=0}^{s-1}C_{N-\sum_{l=0}^{k-1}n_{ll}}^{n_{kk}}$
is defined with the binomial function $C_{n}^{m}=n!/\left[m!\left(n-m\right)!\right]$.
This expression allows us to rewrite the observable as 
\begin{equation}
\left\langle O\right\rangle =\sum_{\left\{ n_{ll}\right\} }C_{\left\{ n_{ll}\right\} }\left\langle O\left\lceil \left\{ n_{ll}\right\} \right\rfloor \right\rangle .\label{eq:observables}
\end{equation}
Setting $O=\sigma_{ll}$ in Eq. (\ref{eq:observables}) and using Eq. (\ref{eq:ll-left}  we obtain
the population of the $\left|l\right\rangle $ states $P_{l}\equiv\sum_{\left\{ n_{ll}\right\} }C_{\left\{ n_{ll}\right\} }n_{ll}\left\langle \left\{ n_{ll}\right\} \right\rangle $.
Setting $O=\sigma_{ll'}$ and using Eq. (\ref{eq:ll'-left})
we obtain the polarization $C_{ll'}$ of the $l-l'$
transition $C_{ll'}\equiv\sum_{\left\{ n_{ll}\right\} }C_{\left\{ n_{ll}\right\} }n_{l'l'}\left\langle \left\{ n_{l'l'}-1,n_{ll'}+1\right\} \right\rangle .$

If we drive the atoms coherently or pump them incoherently, we 
arrive at steady-state. In this case, we can calculate   
the steady-state spectrum of the atoms $S\left(\omega\right)$. 
 According to quantum regression theorem \citep{PMeystre},
we have $S\left(\omega\right)\propto\sum_{l>l'}\Gamma_{ll'}\textrm{Re}\int_{0}^{\infty}d\tau e^{-i\omega\tau}\mathrm{tr}\left\{ \sigma_{ll'}\tilde{\rho}\left(\tau\right)\right\} $
with the operator $\tilde{\rho}\left(\tau\right)$, which satisfies
 the master equation as $\rho$ with however the initial condition
$\sigma_{l'l}\rho_{std}$, where $\rho_{std}$ is the density operator
at the steady-state. Furthermore, we can utilize Eq. (\ref{eq:identity})
to rewrite the expression inside the integral as $\mathrm{tr}\left\{ \sigma_{ll'}\tilde{\rho}\left(\tau\right)\right\} =\sum_{\left\{ n_{kk}\right\} }C_{\left\{ n_{kk}\right\} }n_{l'k}\left\langle \left\{ n_{kk}-1,n_{lk}=1\right\} \right\rangle _{\tilde{\rho}\left(\tau\right)}.$
Here, we have used Eq. (\ref{eq:ll'-left}) and introduced the function
$\left\langle n\right\rangle _{\tilde{\rho}\left(\tau\right)}=\mathrm{tr}\left\{ \left\lceil n\right\rfloor \tilde{\rho}\left(\tau\right)\right\} $.
The term on the right side differs from $\left\langle \left\{ n_{kk}\right\} \right\rangle _{\tilde{\rho}\left(\tau\right)}$
by that $n_{kk}$ reduces by one and $n_{lk}$ becomes one. The function
$\left\langle n\right\rangle _{\tilde{\rho}\left(\tau\right)}$ satisfies
the same equation as $\left\langle n\right\rangle $ with however
the initial condition $\left\langle n\right\rangle _{\tilde{\rho}\left(0\right)}=\sum_{k}n_{kl'}\left\langle \left\{ n_{kl'}-1,n_{kl}+1\right\} \right\rangle _{std}$.
We obtain this condition with the help of Eq. (\ref{eq:ll'-right}).
By integrating the spectrum with respect to the frequency, we 
obtain the radiation intensity $I=I_{ind}+I_{c}$, which includes
the contribution from individual atoms $I_{ind}\propto\sum_{l>l'}\Gamma_{ll'}C_{\left\{ n_{ll}\right\} }n_{ll}\left\langle \left\{ n_{ll}\right\} \right\rangle _{std}$
and from the atomic correlation $I_{c}\propto\sum_{l>l'}\Gamma_{ll'}C_{\left\{ n_{ll}\right\} }n_{ll}n_{l'l'}\left\langle n_{ll}-1,n_{l'l'}-1,n_{l'l}=1,n_{l'l'}=1\right\rangle _{std}$.
Here, we have utilized Eq. (\ref{eq:ll'-left-left}) and (\ref{eq:identity}).
Notice that $I_{c}$ is determined by the off-diagonal elements
of $\left\langle n\right\rangle $. 

\subsection{Numerical Implementation and Complexity Analysis \label{subsec:numerical-solution-complexity}}

In the following, we discuss how to solve the equation for $\left\langle n\right\rangle $
numerically. To this end, we should first find an efficient
way of representing $\left\langle n\right\rangle $ in the computer.
In principle, we can view $\left\langle n\right\rangle $ as a matrix
with $s^{2}$ dimensions and $N$ elements for each dimension. In
this case, the matrix  is a very sparse because the elements satisfying
$\sum_{\left\{ n_{ll'}\right\} }n_{ll'}\neq N$ are not necessary.
In practice, we can define an one-dimensional array to represent
only the $N_{dm}$ (specified later on) necessary elements $\left\langle n\right\rangle $.
Then, the remaining question is how to relate the index of elements $i$ with the set
of numbers $\left\{ n_{ll'}\right\} $. The easy solution is to define
a $N_{dm}\times s^{2}$ two-dimensional ancillary array, where the
first dimension specifies $i$ and the second dimension the values
$\left\{ n_{ll'}\right\} $. To reduce the size of the ancillary array,
we need a clever way to compress this array. We propose to compress
the set $\left\{ n_{ll'}\right\} $ to single number $n_{i}$ with
the relation $n_{i}=\sum_{l,l'=0}^{s-1}\left(N^{s\times l+l'}+n_{ll'}\right)$.
Then, we can use the index $i$ of $\left\langle n\right\rangle $
to retrieve $n_{i}$ from the ancillary array and then obtain $n_{ll'}$
by dividing $n_{i}$ sequentially with the power of $N$ and taking
the residual. In return, if we know $n_{ll'}$, we can calculate $n_{i}$
with the expression given above and then locate the position $i$
of $n_{i}$ in the integer array and finally utilize $i$ to find
the required $\left\langle n\right\rangle $. 

The above discussion indicates that the complexity of solving the
equation numerically is mainly determined by the number $N_{dm}$
of the elements $\left\langle n\right\rangle =\mathrm{tr}\left\{ \rho \left\lceil n\right\rfloor \right\} $.
To compute this number, we recall that $\left\lceil n\right\rfloor $
are specified by a set of numbers $\left\{ n_{ll'}\right\} $ under
the restriction $\sum_{l,l'}n_{ll'}=N$. If we image the atoms
as $N$ balls and $n_{ll'}$ the number of balls in the box labeled
by $ll'$, the searched number is equivalent to the number of possibilities
of arranging $N$ balls in the boxes. After some calculation, we get
the number $N_{dm}=C_{N+s^{2}-1}^{s^{2}-1}$ and can approximate it
as $N^{s^{2}-1}$. We see that this number increases polynomially
with the number of atoms $N$ but exponentially with the number $s$
of levels. As an example, the number is estimated as $N^{3},N^{8},N^{15}$
for $N$ two-, three- and four-level atoms, respectively. We estimate
this number as $1.5\times10^{7}$ for $250$ two-, $25$ three- or
$12$ four-level atoms and thus we need  about $5$ Gb memory to represent
$\left\langle n\right\rangle $ in the computer if we assume $2$
bits for one element. This simple calculation establishes the maximum
number of atoms that can be simulated with a normal computer. 

\section{Superradiance of Hundreds of Two-level atoms \label{sec:applications}}

In this section, we apply our collective description of density matrix to two-level atoms by simply
restricting the level indices $l,l'$ to zero (lower level) and one
(upper level) and study the superradiance from hundreds of atoms,
which are either initially excited, coherently driven or incoherently
pumped. To understand the results, we should keep in mind that $\gamma_{10},\gamma_{01},\xi_{10}$
are the decay, pumping and dephasing rate of individual atoms while
$\Gamma_{10}$ is the collective decay rate. 

As explained in the introduction, this system is often described by
the superradiance master equation, which is defined with the collective
Pauli operator $\sigma_{z}=\sum_{j}\sigma_{z}^{j}$ and the collective
creation $\sigma_{\text{+}}=\sum_{j}\sigma_{\text{+}}^{j}$ and annihilation
ladder operator $\sigma_{-}=\sum_{j}\sigma_{-}^{j}$. This description
is equivalent to our description because of $\sigma_{z}^{j}=\sigma_{11}^{j}-\sigma_{00}^{j}$,
$\sigma_{\text{+}}^{j}=\sigma_{10}^{j}$ and $\sigma_{-}^{j}=\sigma_{01}^{j}$.
In addition, the description with $\sigma_{z},\sigma_{\pm}$ also
implies that the two-level atoms are equivalent to spin-1/2 particles.
Thus, we can also introduce the spin operator $j_{x}=\left(\sigma_{-}+\sigma_{+}\right)/2,j_{y}=i\left(\sigma_{-}-\sigma_{+}\right)/2,j_{z}=\sigma_{z}/2$
and visualize the atomic dynamics with angular moment $\mathbf{J}=\sum_{i=x,y,z}J_{i}\mathbf{e}_{i}$
and their uncertainty $\Delta\mathbf{J}=\sum_{i}\Delta J_{i}\mathbf{e}_{i}$,
where $J_{i}=\left\langle j_{i}\right\rangle $ and $\Delta J_{i}=\sqrt{\left\langle j_{i}^{2}\right\rangle -J_{i}^{2}}$
are the components in Canteen coordinate system. The expressions to
compute these components can be found in Appendix \ref{sec:angular-momentum-uncertainty}. 

\subsection{Superradiance from Atoms Initially Excited }

\begin{figure}
\begin{centering}
\includegraphics[scale=0.9]{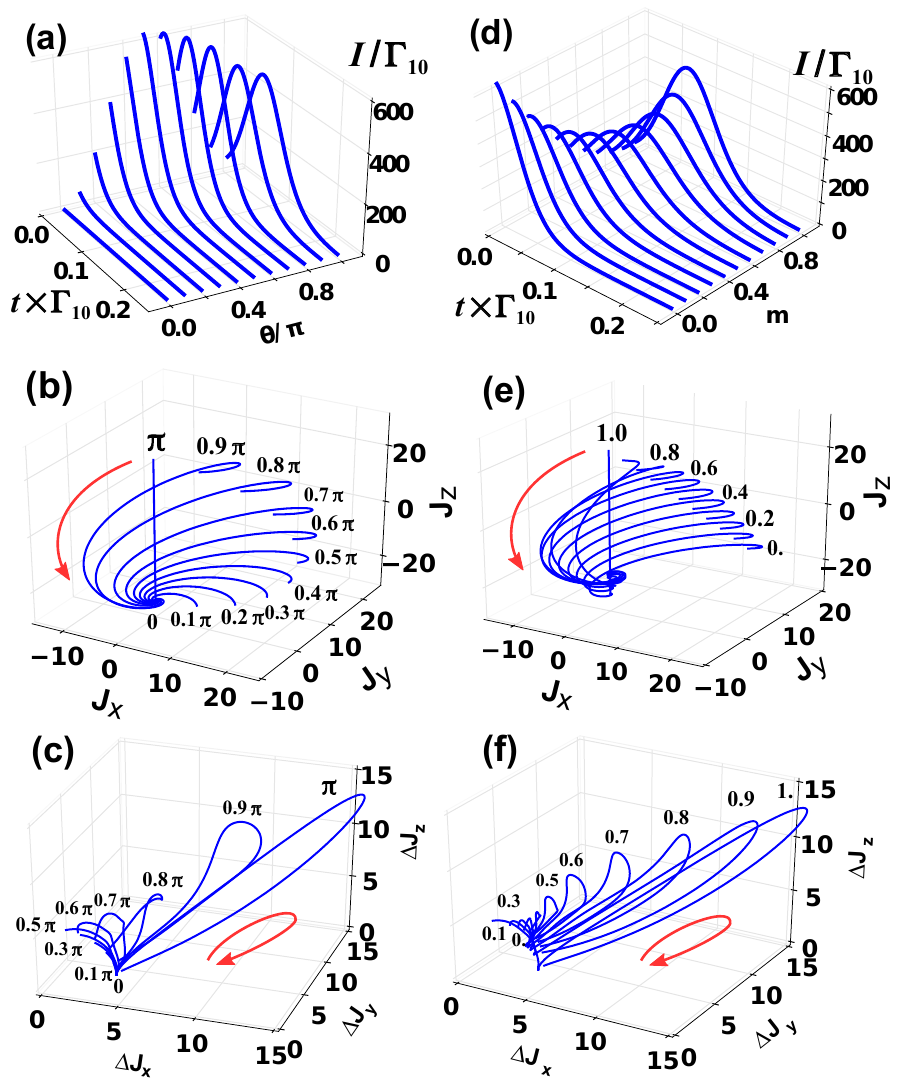}
\par\end{centering}
\centering{}\caption{\label{fig:initial-states}Superradiance (a,d), angular momentum
vector $\mathbf{J}$ (b,d) and uncertainty vector $\Delta\mathbf{J}$
(e,f) for fifty identical two-level atoms. (a-c) show the influence
of the initial pure states with varying angle $\theta_{i}$ but fixed
$\phi_{i}=0$. (d-f) show the influence of increasing mixing (probability
$m$) of two pure states with $\left(\theta_{1},\phi_{1}\right)=\left(0,0\right)$
and $\left(\theta_{2},\phi_{2}\right)=\left(\pi/2,0\right)$. $\omega_{c}=\omega_{1}-\omega_{0}=10\pi\Gamma_{10}$
and other parameters are zero. For more details see text. }
\end{figure}

In this subsection, we consider the superradiance from two-level atoms
which are initially excited. Applying Eq. (\ref{eq:initial-values})
we can specify the initial collective density matrix as $\left\langle n\right\rangle _{0}=\prod_{l,l'=0,1}\left(\sum_{i}m_{i}c_{l}^{i}c_{l'}^{i*}\right)^{n_{ll'}}$.
Here, the complex numbers $c_{1}^{i},c_{2}^{i}$ specify the pure
atomic states and $m_{i}$ specify the probability of these states.
Furthermore, the complex numbers can be parameterized as $c_{1}^{i}=\sin\left(\theta_{i}/2\right)e^{i\phi_{i}}$
and $c_{0}^{i}=\cos\left(\theta_{i}/2\right)$ with one azimuth angle
$\theta_{i}\in\left[0,\pi\right]$ and one polar angle $\phi_{i}\in\left[0,2\pi\right]$
\citep{LMandel}. Fig. \ref{fig:initial-states} shows how the atomic
initial states affect the superradiance and the atomic dynamics for
fifty identical two-level atoms. We assume that the atoms are resonant
to the cavity mode, i.e. $\omega_{1}-\omega_{0}=\omega_{c}$, which
results to vanishing Lamb shift, i.e. $\Omega_{ll'}=0$. For better
visualization we use small atomic transition frequency $\omega_{1}-\omega_{0}=10\pi\Gamma_{10}$.

Fig. \ref{fig:initial-states} (a,b,c) show the results when the initial
states are pure states. Fig. \ref{fig:initial-states} (a) shows that
the initial radiation rate is finite for $\theta_{i}=\pi$ (the fully
excited state), and increases to the maximum for $\theta_{i}=\pi/2$
(the superposition state), and decreases back to zero for $\theta_{i}=0$
(the ground state). The radiation has a pulse structure for $\theta_{i}\in\left(\pi/2,\pi\right]$
and the pulse maximum increases and the pulse center shifts to the
earlier time with reducing $\theta_{i}$. In contrast, the radiation
shows an exponential decay for $\theta_{i}\in\left(0,\pi/2\right]$
and the time when the radiation vanishes reduces with reducing $\theta_{i}$.
Fig. \ref{fig:initial-states} (b) shows that the initial angular
momentum $\mathbf{J}$ rotates from the north pole to the equator,
and finally to the north pole. In particular, for $\theta_{i}=\pi$
the components $J_{x},J_{y}$ are zero. That is to say, the superradiance
from the atoms fully excited is not accompanied by the formation of
dipole. Actually, this is the superradiance initially discussed by Dicke
\citep{RHDicke}. However, for $\theta_{i}\neq0$ the vector $\mathbf{J}$
actually circles around the z-axis before ending up at the south pole.
Notice that the components $J_{x},J_{y}$ are not zero. That is to
say that for the atoms, which are not fully excited, the superradiance
is accompanied by the formation of macroscopic dipole. This corresponds actually
to the so-called superfluorescence \citep{JCMacGillivray}. Fig. \ref{fig:initial-states}
(c) shows that the initial uncertainty of angular momentum $\Delta\mathbf{J}$
vector has a fixed length \footnote{The product states of two-level systems are also known as spin coherent
states. For these states the angular momentum uncertainties are $\Delta J_{i}=\sqrt{N/4-N\left\langle j_{i}\right\rangle ^{2}}$
with a maximum $\sqrt{N/4}$ \citep{JMa}. } of $\sqrt{N/4}\approx3.54$ and points to different directions in
the plane with $\Delta J_{y}=3.54$ for different $\theta_{i}$. For
$\theta_{i}\in\left[\pi/2,\pi\right]$ the length of that vector increases
with time and the maximal length reduces with reducing $\theta_{i}$.
In contrast, for $\theta_{i}\in\left[0,\pi/2\right]$ the length does
not change with time. In all the cases the $\Delta\mathbf{J}$ vector
ends at the point $\left(3.54,3.54,0\right)$. In addition, we have
also varied the angle $\phi_{i}$ but found no influence on the superradiance. 

Fig. \ref{fig:initial-states} (d,e,f) show how the mixed initial
states affect the superradiance and the atomic dynamics. In particular,
we consider that every single atom is initially in a mixture of the
excited state $\left(\theta_{1},\phi_{1}\right)=\left(0,0\right)$
with a probability $m$ and the superposition state $\left(\theta_{2},\phi_{2}\right)=\left(\pi/2,0\right)$
with a probability $1-m$. Fig. \ref{fig:initial-states} (d) shows
that the initial radiation rate reduces with increasing $m$ due to
the reduced contribution of the superposition state while the radiation
maximum increases because of increased contribution of the excited
state. Fig. \ref{fig:initial-states} (e) shows that the initial $\mathbf{J}$
vector moves from the equator to the north pole with increasing $m$.
$\mathbf{J}$ circles around the z-axis for $m\neq1$ but shrinks
along the $z$-axis for $m=0$. In both cases $\mathbf{J}$ ends eventually
at the south pole. Fig. \ref{fig:initial-states} (f) shows that the
initial $\Delta\mathbf{J}$ vector goes upwards when $m$ increases
from $0$ to $0.5$ but then goes downwards when $m$ further increases
to one. The length of the vector increases firstly and then reduces
with increasing time. $\Delta\mathbf{J}$ always ends up at the point
$\left(3.54,3.54,0\right)$ corresponding to the fully ground state.
Notice that the maximal length increases with increasing $m$. 

In Appendix \ref{subsec:influence-superradiance-pulse} we have further
studied how the superradiance and the atomic dynamics are affected
by the number of atoms $N$, the frequency detuning $\chi_{10}=\omega_{1}-\omega_{0}-\omega_{c}$,
the decay rate $\gamma_{10}$ and the dephasing rate $\xi_{10}$ of
individual atoms. We find that the maximum, center, duration of the
superradiance pulses change with $N$ according to $N^{2},N^{-1}\mathrm{ln}N,N^{-1}$,
respectively. The maximum depends on $\chi_{10}$ according to a Lorentzian
shape while the center and duration to a parabolic shape. The maximum
changes with $\gamma_{10}$ according to $\exp\left(-\alpha\gamma_{10}\right)$
while the center, duration according to $\left[\gamma_{10}^{2}+\beta\right]^{-2}$
($\alpha,\beta$ are some constant), respectively. 

\subsection{Superradiance from Atoms Coherently Driven \label{subsec:Contineous-Superradiance-Atoms-Coherently-Pumping}}

\begin{figure}
\begin{centering}
\includegraphics[scale=0.2]{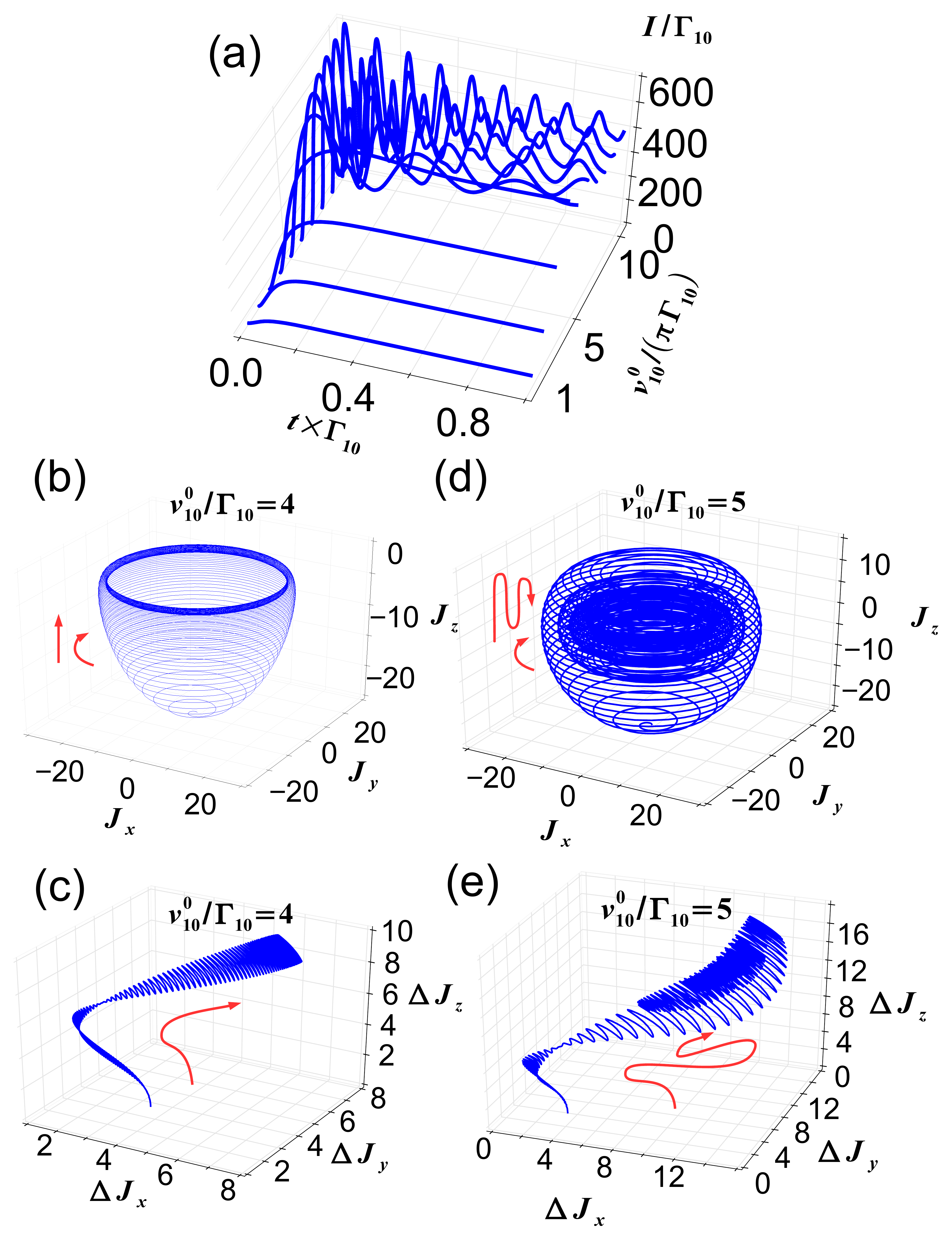}
\par\end{centering}
\caption{\label{fig:Superradiance-coherent-pumping}Superradiance (a) and dynamics
of angular momentum vector $\mathbf{J}$ (b,d) and its uncertainty
$\Delta\mathbf{J}$ (c,e) for fifty identical two-level atoms, which
are initially on the ground state and are then driven coherently with
a strength $v_{10}^{0}$. (b,c) are results under weak driving $v_{10}^{0}/\left(\pi\Gamma_{10}\right)=4$
while (d,e) under moderate driving $v_{10}^{0}/\left(\pi\Gamma_{10}\right)=5$.
We assume that the atoms are resonant to the cavity mode and also to the
driving laser, i.e. $\omega_{1}-\omega_{0}=\omega_{d}=\omega_{c}$.
We use small transition frequency $\omega_{1}-\omega_{0}=200\times\pi\Gamma_{10}$
to better illustrate the dynamics. The red curves with arrows indicate
the direction of the dynamics. Other parameters are zero.
For more details see text. }
\end{figure}

In this subsection, we analyze the superradiance and the dynamics
of fifty identical two-level atoms, see Fig. \ref{fig:Superradiance-coherent-pumping},
which are driven coherently by an external field with a strength $v_{10}^{0}$
and experience simultaneously a collective decay with the rate $\Gamma_{10}$.
If the atoms are only driven coherently, their population follows
Rabi oscillation with a period $T=\pi/v_{10}^{0}$. If the atoms experience
only the collective decay, their population on the excited state decays
in time and the time for vanishing population depends on the number
of atoms. For fifty atoms as considered here, we estimate this time as
$\tau_{a}=0.2/\Gamma_{10}$. 

With the parameters as analyzed above we can now understand the results shown in
Fig. \ref{fig:Superradiance-coherent-pumping}. For $\tau_{a}/T<1$
or $v_{10}^{0}/\pi<\Gamma_{10}$, the system achieves the steady-state
before it starts oscillating because the collective decay is relatively
faster than the coherent driving. This leads to the smooth change
of the radiation for $v_{10}^{0}/\left(\pi\Gamma_{10}\right)<4$ as shown
in Fig. \ref{fig:Superradiance-coherent-pumping} (a), the cup-like
trajectory of the vector $\mathbf{J}$ and the hoe-like trajectory
of the vector $\Delta\mathbf{J}$ as shown in Fig. \ref{fig:Superradiance-coherent-pumping}
(b,c), respectively. Here, the oscillation frequency of $J_{x}$,$J_{y}$
and $\Delta J_{x}$, $\Delta J_{y}$ is determined by the frequency
of the external field. By increasing the driving, the ratio $v_{10}^{0}/\left(\pi\Gamma_{10}\right)$
increases. This leads to the increased steady-state radiation, see 
Fig. \ref{fig:Superradiance-coherent-pumping} (a), and the increased
height of the cup-like trajectory of $\mathbf{J}$ as well as the
increased length of the hoe-like trajectory of $\Delta\mathbf{J}$
(not shown). 

For $\tau_{a}/T=1$, i.e. $v_{10}^{0}/\left(\pi\Gamma_{10}\right)=5$,
the two processes are comparable and the system shows oscillations
before achieving the steady state, see Fig. \ref{fig:Superradiance-coherent-pumping}
(a), which is accompanied by the vase-like trajectory of $\mathbf{J}$
and the tie-like trajectory of $\Delta\mathbf{J}$ as shown in Fig. \ref{fig:Superradiance-coherent-pumping}(d,e),
respectively. We notice that each oscillation of the radiation is
related to one shell of the vase-like trajectory and one fold of the
tie-like trajectory. The reduced amplitude of radiation oscillation
is related to reduced radius of the shell and of the folded layer.
In addition, for $\tau_{a}/T>1$, i.e. $v_{10}^{0}/\left(\pi\Gamma_{10}\right)>5$,
the coherent driving is much faster than the collective decay. As  a result,
we see more oscillations with reduced period  before the system achieves
the steady-state, see Fig. \ref{fig:Superradiance-coherent-pumping}(a),
which is accompanied by the spherical lantern-like trajectory of $\mathbf{J}$
and the banana-like trajectory of $\Delta\mathbf{J}$ (not shown),
respectively. At this point, we might conclude that the number of
turns in the outer shell reduces with increasing driving of the external
field. 

In Appendix \ref{subsec:influence-superradiance-driven} we have further
studied how the superradiance from the atoms driven coherently is
influenced by the number of atoms $N$, the frequency detuning
$\chi_{10}=\omega_{1}-\omega_{0}-\omega_{c}$ , the decay rate $\gamma_{10}$
and dephasing rate $\xi_{10}$ of individual atoms. We find that with
increasing $N$ the superradiance changes from the oscillatory behavior
to the steady behavior at long time (due to enhanced collective decay
rate). With increasing $\chi_{10}$ the superradiance maximum reduces and
the time to achieve steady superradiance reduces (due to off-resonant
excitation). With increasing $\gamma_{10},\xi_{10}$, the time to achieve
steady radiation reduces and the steady radiation also reduces. In
general, $\gamma_{10}$ affects the superradiance much stronger than
$\xi_{10}$ since it affects also the population. 

\subsection{Superradiance from Atoms Pumped Incoherently \label{sec:incoherent-individual-pumping}}

\begin{figure}
\begin{centering}
\includegraphics[scale=0.9]{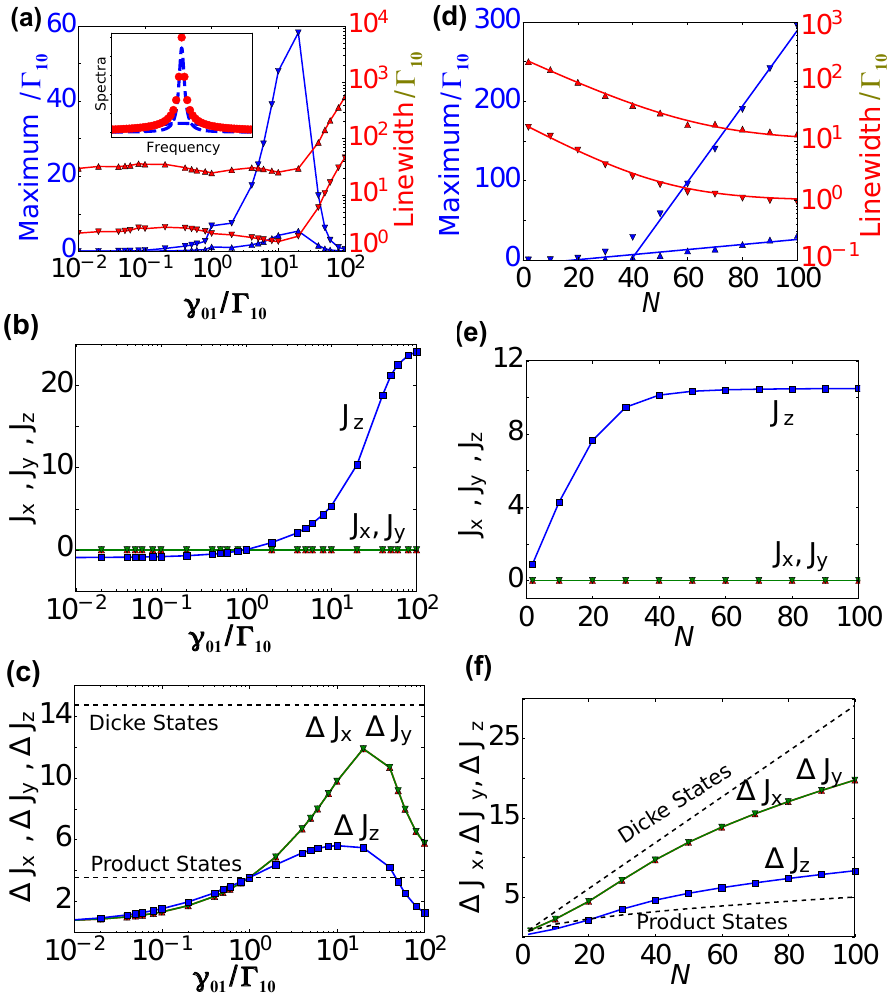}
\par\end{centering}
\caption{\label{fig:incoherent-individual-pumping} Steady-state spectra, consisting
of a sharp peak and a broad background {[}see inset of (a){]}, for
increasing pumping rate $\gamma_{01}$ (a,b,c, for fifty atoms) and
number of atoms $N$ (d,e,f, for fixed pumping rate $\gamma_{01}=20\gamma_{10}$).
In (a,d) we show the maximum (blue triangles) and linewidth (red triangles)
of the sharp peak (lower triangles) and of the background (upper triangles).
In (d) the maximum and the linewidth of the sharp peak (background)
are fitted with $-188.05+4.77N$ ($-6.01+0.32N$) and $1.0+18.3\exp\left(-5.7\times10^{-2}N\right)$
{[}$10.1+230\exp\left(-4.9\times10^{-2}N\right)${]}. (b-f) show the
angular momentum components and their uncertainties: $J_{x}$ and
uncertainty $\Delta J_{x}$ (red upper triangles), $J_{y}$ and $\Delta J_{y}$
(green lower triangles), $J_{z}$ and $\Delta J_{z}$ (blue squares).
In (c,f) the upper and lower dashed line indicate the limit of the
Dicke superradiant states and the atomic product states , respectively.
Other parameters are zero. For more details see text. }
\end{figure}

Steady-state superradiance was firstly studied by D. Meiser and M.
J. Holland \citep{DMeiser,DMeiser1} by solving the superradiance
master equation with Monte-Carlo method for tens of atoms \citep{DMeiser1}
and second-order mean field theory \citep{DMeiser} for thousands
of atoms. Here, we apply our description to study this phenomenon
and unravel more details with exact simulations, see Fig. \ref{fig:incoherent-individual-pumping}. 

Fig. \ref{fig:incoherent-individual-pumping} (a,b,c) show the steady-state
spectra and the steady-states of fifty identical two-level
atoms for increasing pumping rate $\gamma_{01}$. The steady-state
spectra consist of not only one sharp peak as discussed in \citep{DMeiser1}
but also one broad background, see inset of Fig. \ref{fig:incoherent-individual-pumping}(a).
Fitting the spectra with two Lorentzian functions, we have determined
the maximum and the linewidth of both the sharp peak and the background,
see Fig. \ref{fig:incoherent-individual-pumping} (a). In general,
the background is about one order of magnitude weaker and broader
than the sharp peak. We see that the maximum (blue lines) increases
slowly with increasing $\gamma_{01}$ for $\gamma_{01}<\Gamma_{10}$
and then quickly for $\gamma_{01}>\gamma_{c}$, but finally decreases
for much larger $\gamma_{01}$. The linewidth (red lines) behaves
in opposite to the maximum, which is a reminiscence of Schawlow-Townes
relation \citep{SchawlowTownes}. The smallest linewidth of the sharp
peak approaches the collective decay rate $\Gamma_{10}$, which agrees
with the results in \citep{DMeiser,DMeiser1}. 

We have also computed the total radiation rate $I_{tot}$ and decompose
it into the contribution from individual atoms $I_{ind}$ and from
the atom-atom correlation $I_{col}$. We find that $I_{tot}$ changes similarly
as the spectrum maximum. Importantly, for $\gamma_{01}<\Gamma_{10}$,
$I_{col}$ is negative and reduces $I_{tot}$. Thus, we can identify
this regime as subradiance regime. For $\gamma_{01}>\Gamma_{10}$, $I_{c}$
is positive and dominates over $I_{ind}$. We can identify this regime
as superradiance regime. For $\gamma_{01}\gg\Gamma_{10}$, $I_{col}$ reduces
and thus $I_{tot}$ also reduces. This reduction can be attributed
to the reduced coupling for higher populated atoms \citep{KDebnath}.
We can identify this regime as the (population) saturation regime
(see below). 

Fig. \ref{fig:incoherent-individual-pumping} (b) shows that the three
regimes as identified before are correlated with $J_{z}$ (blue lines),
which increases from negative value to positive value (inverted population)
and finally saturated value, respectively. Fig. \ref{fig:incoherent-individual-pumping}
(c) shows a much close correlation with the
uncertainties $\Delta J_{x},\Delta J_{y},\Delta J_{z}$. In particular,
the uncertainties are smaller than the limit $\sqrt{N/4}=3.5$ (the
uncertainty of uncorrelated atoms) for $\gamma_{01}<\Gamma_{10}$,
but become larger than that value for $\gamma_{01}>\Gamma_{10}$. In addition,
they approach the limit $\sqrt{\left(N/2\right)\left(N/2+1\right)/3}=15.3$
(the uncertainty of correlated atoms in Dicke superradiant states
\footnote{If the atoms are in Dicke states $\left|J,M\right\rangle $, we have
$\left\langle j_{x}\right\rangle =\left\langle j_{y}\right\rangle =0$
and $\left\langle j_{z}\right\rangle =M$ as well as $\left\langle j^{2}\right\rangle =J\left(J+1\right)$.
Thus, we have $\Delta J_{x}=\sqrt{\left\langle j_{x}^{2}\right\rangle }$,
$\Delta J_{y}=\sqrt{\left\langle j_{y}^{2}\right\rangle }$ and $\Delta J_{z}=\sqrt{\left\langle j_{z}^{2}\right\rangle -M}$.
If we assume $\left\langle j_{x}^{2}\right\rangle =\left\langle j_{y}^{2}\right\rangle =\left\langle j_{z}^{2}\right\rangle $
we  have $\Delta J_{x}=\Delta J_{y}=\sqrt{J\left(J+1\right)/3}$,
and the maximum $\Delta J_{z}=\sqrt{J\left(J+1\right)/3}$ (for $M=0$).
For the Dicke superradiant states, we have $J=N/2$ and thus the limit
$\sqrt{\left(N/2\right)\left(N/2+1\right)/3}$ as discussed in the
main text. }) for larger $\gamma_{01}$. Notice that $J_{x}$ and $J_{y}$ are
always zero and thus there is no macroscopic dipole involved. 

Fig. \ref{fig:incoherent-individual-pumping} (d-f) show the steady-state
spectra and the atomic steady-states for increasing number of atoms
$N$. Fig. \ref{fig:incoherent-individual-pumping} (d) shows that
the maximum of the sharp peak and the background increases quadratically
with $N$ for $N<40$ but linearly for $N>40$ while the linewidth
reduces exponentially for small $N$ but approaches a constant for
large $N$. Fig. \ref{fig:incoherent-individual-pumping} (e) shows
that the change of the peak maximum is correlated with $J_{z}$, which
increases initially but saturates for large $N$. Fig. \ref{fig:incoherent-individual-pumping}(f)
shows that the maximum is more closely correlated with the uncertainty
$\Delta J_{l}$. Moreover, we find that $\Delta J_{z}$ departs from
the limit $\sqrt{N/4}$ while $\Delta J_{x},\Delta J_{y}$ from $\sqrt{\left(N/2\right)\left(N/2+1\right)/3}$. 
This indicates that the atoms are between the uncorrelated state and the fully correclated state.

In Appendix \ref{subsec:influence-superradiance-driven}, we have
further studied how the steady-state spectrum is influenced by the
decay rate $\gamma_{10}$ and the dephasing rate $\xi_{10}$ of individual
atoms. We find that $\gamma_{10}$ affects the spectrum much stronger
than $\xi_{10}$ because $\gamma_{10}$ competes directly with the
incoherently pumping $\gamma_{01}$. Increasing $\gamma_{10}$ just
tens times, we reduce the spectrum maximum and increase the linewidth
by orders of magnitude. In contrast, only by increasing $\xi_{10}$
by one hundred times, we can only reduce the spectrum maximum and
linewidth by two or three times. These results suggest that the steady-state
superradiance seem to be  robust to the dephasing of individual atoms  but
sensitive to the decay of individual atoms. 

\section{Conclusions \label{sec:Conclusions}}

To sum up, in this article, we presented a collective description
of density matrix for identical multi-level atoms. Our description
explores symmetry of density matrix in the basis of atomic product
states without explicitly applying the symmetry group theory. Since
our description removes the redundancy of density matrix elements,
it is possible to carry out exact simulation for systems with hundreds
of two-level atoms and tens of three- or four-level atoms. 

As an example, we applied our description to simulate superradiance
from hundreds of two-level atoms, which are either excited initially
or driven coherently or pumped incoherently. With the simulation we
identified two kinds of superradiance. The first kind does not involve
the formation of dipole and appears for the atoms, which are either
fully excited initially or pumped incoherently. The second kind
does involve the dipole and appears for the atoms, which are initially
in superposition states or driven coherently. In addition, we found
that the superradiance is more closely correlated with the uncertainties
of angular momentum. 

In the future, we can apply our description to study superradiance
from multi-level atoms \citep{RTSutherland}. In addition, we can
also extend our description in many directions. By extending the description
to atoms of multi-species, we can study the cavity-mediated interaction
between the species and the resulting effects, such as superradiance
beats \citep{MANorcia}, phase synchronization \citep{MXu-1,JMWeiner}.
By including the cavity mode directly rather than eliminating it,
we can also develop a description to study the effects in the intermediate
or good cavity limit, such as superradiance-to-lasing transition \citep{KDebnath,MANorcia-1,DATieri}.
By including measurement backactions we can also study the measurement
and control of quantum system \citep{HMWiseman}, e.g. the conditional
spin-squeezing \citep{ZChen}. 
\begin{acknowledgments}
Y. Z. acknowledges Klaus M{\o}lmer, Jiabao Chen and Shi-Lei Su for several
illuminating discussions. This work was supported by Villum Foundation
(Y. Z.). 
\end{acknowledgments}

\appendix

\section{Supplemental Results\label{sec:Supplemential-Material}}

In the main text, we have presented the main results on the superradiance
from hundreds of two-level atoms, which are either excited initially
or driven coherently or pumped incoherently. In this Appendix, we
supplement these results by analyzing the influence of the number
of atoms, the decay and dephasing of individual atoms. 

\subsection{Influence on Superradiance from Atom Initially Excited \label{subsec:influence-superradiance-pulse}}

\begin{figure}
\begin{centering}
\includegraphics{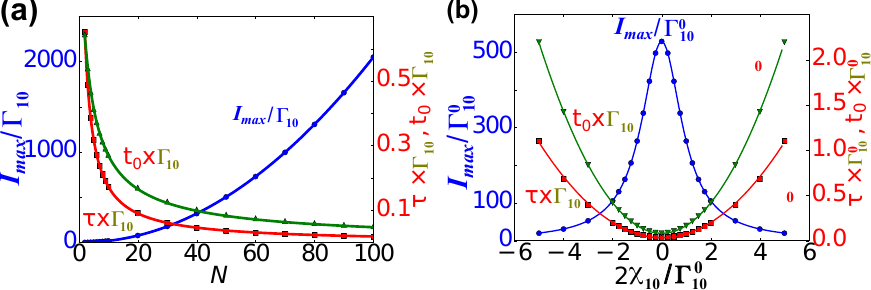}
\par\end{centering}
\begin{centering}
\includegraphics{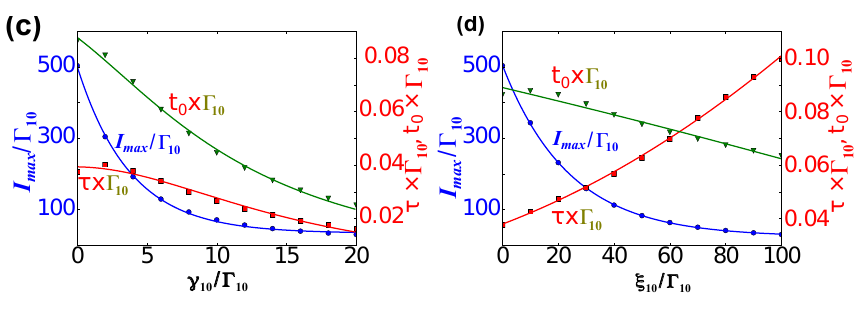}
\par\end{centering}
\centering{}\caption{\label{fig:superradiance} Maximum $I_{max}$ (dots, blue curves),
center $t_{0}$ ( triangles, green curves) and duration $\tau$ (squares,
red curves) of superradiance pulses for increasing number of atoms
$N$ (a), frequency detuning $\chi_{10}=\omega_{1}-\omega_{0}-\omega_{c}$
(b), decay rate $\gamma_{10}$ (c) and dephasing rate $\xi_{10}$
(d). In (b) $\Gamma_{10}^{0}$ is the collective decay rate on resonant
condition. In (a,c,d) we focus on the resonant case such that $\Gamma_{10}=\Gamma_{10}^{0}$.
In (a) the data are fitted with $I_{max}/\Gamma_{10}=0.54-0.37N+0.21N^{2}$,
$t_{0}\Gamma_{10}=\left(0.88/N\right)\ln\left(2.12N\right)$, $\tau\Gamma_{10}=1.88/(0.87+N)$.
In (b) the data are fitted with $I_{max}/\Gamma_{10}^{0}=529\left[\left(\alpha-1.38\right){}^{2}+1.\right]^{-1}$,
$t_{0}\Gamma_{10}^{0}=0.03-2.18\alpha+0.04\alpha^{2}$, $\tau\Gamma_{10}^{0}=0.08-2.18\alpha+0.08\alpha^{2}$
with $\alpha=2\chi/\Gamma_{10}^{0}$. In (c) the data are fitted with
$I_{max}/\Gamma_{10}=33.84+466.99\exp\left(-0.27\beta\right)$, $t_{0}\Gamma_{10}=0.10\times13.89^{2}\left[\left(\beta+4.89\right)^{2}+14^{2}\right]^{-1}$,
$\tau\Gamma_{10}=0.10\times15.83^{2}\left[\left(\beta+0.26\right)^{2}+16{}^{2}\right]^{-1}$
with the ratio $\beta=\gamma_{10}/\gamma_{c}$. In (d) the data are
fitted with $I_{max}/\Gamma_{10}=24+480.5\exp\left(-0.04\delta\right)$,
$t_{0}\Gamma_{10}=8.9\times10^{-2}-2.5\times10^{-4}\delta$, $\tau\Gamma_{10}=3.7\times10^{-2}-3.9\times10^{-4}\delta$
with the ratio $\delta=\xi_{10}/\Gamma_{10}$. Other parameters are
zero. For more details see text. }
\end{figure}

In the following, we discuss the supplemental results on the superradiance
from the atoms, which are initially excited. Fig. \ref{fig:superradiance}
shows how the maximum $I_{max}$, center $t_{0}$ and duration $\tau$
of the superradiant pulses are affected by the number of atoms $N$
(a), the frequency detuning $\chi_{10}=\omega_{1}-\omega_{0}-\omega_{c}$
(b), the decay rate $\gamma_{10}$ (c) and the dephasing rate $\xi_{10}$
(d). Fig. \ref{fig:superradiance} (a) shows that $I_{max}$
increases as $N^{2}$, while $\tau$ and $t_{0}$ vary according
to $1/N$ and $N\ln N$, respectively, which agrees with \citep{LMandel}
except for small deviation for small $N$. We also observe that $J_{z}$
drops from $N/2$ to $-N/2$ with an accelerated speed for large $N$
and the other components vanish, i.e. $J_{x}=J_{y}=0$ (not shown).
In addition, $\Delta J_{z}$, $\Delta J_{x}$, $\Delta J_{y}$ have
same structure as the superradiance pulses and their maximum increase
with $N$ (not shown).

To better interpret Fig. \ref{fig:superradiance} (b), we introduce
the ratio $\alpha=2\chi_{10}/\Gamma_{10}^{0}$ with the collective decay
rate $\Gamma_{10}^{0}$ in the resonant case ($\chi_{10}=0$) and express
the collective decay $\Gamma_{10}\left(\alpha\right)=\Gamma_{10}^{0}/\left(\alpha^{2}+1\right)$
and the Lamb shift $\Omega_{ll'}\left(\alpha\right)=-\Gamma_{10}^{0}\left[\alpha/\left(\alpha^{2}+1\right)\right]$
as functions of this ratio. Fig. \ref{fig:superradiance} (b) shows
that $I_{max}$ reduces with increasing $\left|\alpha\right|$ while
$t_{0}$ and $\tau$ increases. The relation of $I_{max}$ and $\alpha$
can be fitted with a Lorentzian function, and the relation of $t_{0}$,
$\tau$ and $\alpha$ can be well fitted with a parabolic function.
We also find that $J_{x}$, $J_{y}$ vanish and $J_{z}$ drops quickly
for small $\left|\alpha\right|$ but slowly for large $\left|\alpha\right|$
(not shown). In addition, $\Delta J_{z}$ changes similarly as the
radiation.

Fig. \ref{fig:superradiance} (c) shows that $I_{max}$ reduces exponentially
with increasing $\beta\equiv\gamma_{10}/\Gamma_{10}$, while $t_{0}$,
$\tau$ reduce inversely with the square of $\beta$. We also find
that $J_{z}$ follows a $S$-shape for small $\beta$ because of the
dominance of $\gamma_{10}$, but decays exponentially for large $\beta$
due to the dominance of $\Gamma_{10}$ (not shown). In addition, $\Delta J_{z}$
follows the change of the superradiance pulses. Fig. \ref{fig:superradiance}
(d) shows that $I_{max}$ decreases exponentially, $\tau$ reduces linearly,
and $t_{0}$ increases with increasing $\delta=\xi_{10}/\Gamma_{10}$.
We also find that $J_{z}$ drops quickly and follows a S-shape for
small $\delta$. In addition, $\Delta J_{z}$ follows the change of
the superradiance pulses. These results suggest that the decay rate
$\gamma_{10}$ affects the superradaince much stronger than the dephasing
rate $\xi_{10}$ because it affects directly the population.

\subsection{Influence on Superradiance from Atoms Driven Coherently  \label{subsec:influence-superradiance-driven}}

\begin{figure}
\begin{centering}
\includegraphics[scale=0.22]{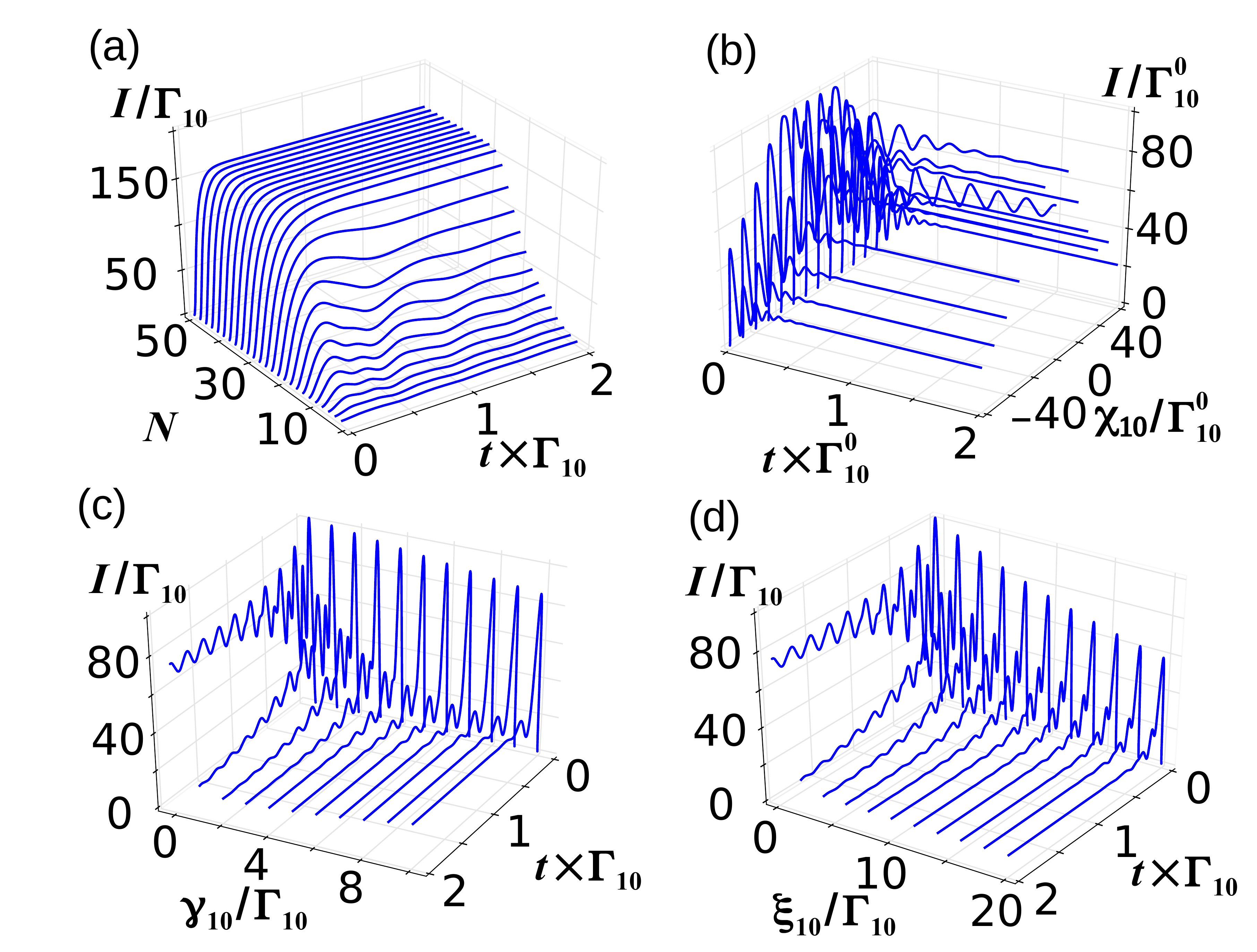}
\par\end{centering}
\caption{\label{fig:influence-driven}Superradiance for increasing number of
atoms $N$ (a), frequency detuning $\chi_{10}=\omega_{1}-\omega_{0}-\omega_{c}$
(b), decay rate $\gamma_{10}$ (c) and dephasing rate $\xi_{10}$
(d). We consider $N=20$ atoms in (b,c,d). The atom-external field
coupling is $v_{10}^{0}/\left(\pi\Gamma_{10}\right)=2$. In (b) $\Gamma_{10}^{0}$
is the collective decay rate $\Gamma_{10}$ on resonant condition
$\chi_{10}=0$. In (a,c,d) we focus on the resonant condition such that
$\Gamma_{10}=\Gamma_{10}^{0}.$ Other parameters are zero. For more
details see text. }
\end{figure}

In the following, we discuss the supplemental results on the superradiance
from the atoms, which are driven coherently. Fig. \ref{fig:influence-driven}
(a) shows that the radiation increases and the number of oscillations
reduces when the number of atoms increases to thirty. This is because
the time $\tau_{a}$ of the collective decay reduces gradually and
becomes comparable with the period $T=\pi/v_{10}^{0}$ of the Rabi
oscillations. The radiation becomes constant and the oscillation disappears
when the number of atoms exceeds thirty because the time $\tau_{a}$
is smaller than the period $T$. Fig. \ref{fig:influence-driven}
(b) shows that the maximum and the steady-state value of the radiation
$I/\Gamma_{10}$ reduce by fifty percents when the frequency $\omega_{d}$
of the external field is detuned from the atomic transition frequency
$\omega_{1}-\omega_{0}$ about sixty times ($\delta=60$) of the collective
decay rate.

Fig. \ref{fig:influence-driven} (c) shows that the maximum of radiation
reduces only a little but the number of oscillations reduces a lot
when the individual decay rate increases to ten times ($\beta=10$)
of the collective decay rate. The steady-state radiation arrives at
the minimum when the two rates are comparable ($\beta=1$) but actually
increases a little for large individual decay rate ($\beta>1$). Fig.
\ref{fig:influence-driven} (d) shows that the individual dephasing
rate affects the radiation in a similar way as the individual decay
rate. However, the similar effect is achieved when the former rate
is about two times of the latter rate, i.e. $\chi\approx2\beta$.

\subsection{Influence on Superradiance from Atoms Incoherently Pumped}

\begin{figure}
\begin{centering}
\includegraphics{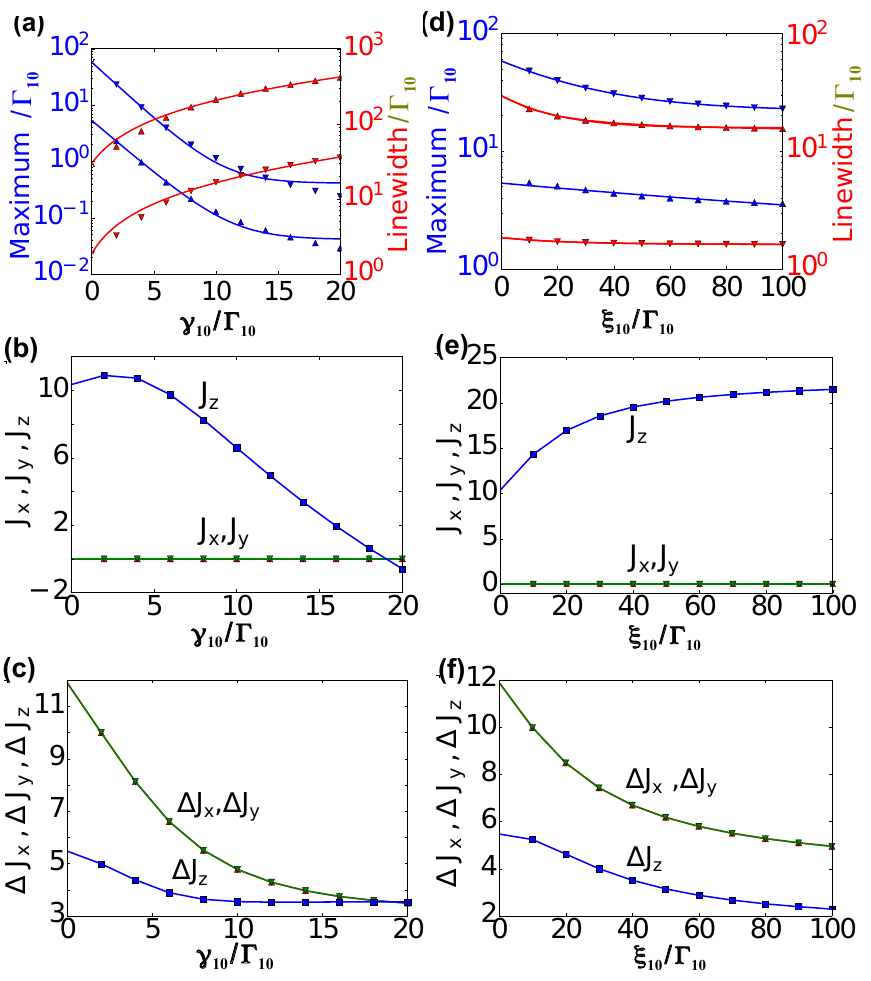}
\par\end{centering}
\caption{\label{fig:individual-decay-dephasing} Influence of decay rate $\gamma_{10}$
(a-c) and dephasing rate $\xi_{10}$ (d-f) on the systems with fifty
atoms. Shown properties are same as in Fig. \ref{fig:incoherent-individual-pumping}.
In (a) the blue upper and lower triangles are fitted with $0.41+57.82\exp\left(-0.47\alpha\right)$,
$58.19\exp\left(-0.43\alpha\right)$ with the ratio $\alpha=\gamma_{10}/\Gamma_{10}$,
respectively, and the red upper and lower triangles are fitted with
$-38.73+40.57\exp\left(3.07\times10^{-2}\alpha\right)$, and $-786.49+788.34\exp\left(1.96\times10^{-2}\alpha\right)$,
respectively. In (d) the blue upper and lower triangles are fitted
with $21.923+36.3\exp\left(-3.53\times10^{-2}\beta\right)$, $1.85+3.51\exp\left(-7.52\times10^{-3}\beta\right)$
with the ratio $\beta=\xi_{10}/\Gamma_{10}$, respectively, and the
red upper and lower triangles are fitted with $1.62+0.23\exp\left(-4.49\times10^{-2}\beta\right)$,
$15.81+13.69\exp\left(6.01\times10^{-2}\beta\right)$, respectively.
$\gamma_{01}=20\Gamma_{10}$ and other parameters are zero. For more
details see text. }
\end{figure}

Fig. \ref{fig:individual-decay-dephasing} (a,b,c) show the influence
of the individual decay rate $\gamma_{10}$ on systems with fifty
atoms. Fig. \ref{fig:individual-decay-dephasing} (a) shows that the
maximum of the peak and background reduces with increasing $\gamma_{10}$,
while their linewidth behaves oppositely. Fig. \ref{fig:individual-decay-dephasing}
(b,c) show that the reduced and broadening spectra are correlated
with the reduced $J_{z}$ and uncertainty $\Delta J_{l}$, respectively.
Notice that $\Delta J_{x}$ and $\Delta J_{y}$ are initially larger
than $\Delta J_{z}$ but approaches it for large $\gamma_{10}$. 

Fig. \ref{fig:individual-decay-dephasing} (d,e,f) show the influence
of the individual dephasing rate $\xi_{10}$. Fig. \ref{fig:individual-decay-dephasing}(d)
shows that the steady-state spectrum behaves differently compared
to Fig. \ref{fig:individual-decay-dephasing} (a) for $\gamma_{10}$.
In this case, the peak linewidth reduces with increasing $\xi_{10}$
while the background linewidth does not change. In addition, the maximum
of peak and background do not reduce so strong as the case for $\gamma_{10}$.
Fig. \ref{fig:individual-decay-dephasing}(e) shows that $J_{z}$
increases first and then saturates with increasing $\xi_{10}$ in
contrast to Fig. \ref{fig:individual-decay-dephasing}(b) for $\gamma_{10}$.
Fig. \ref{fig:individual-decay-dephasing}(f) shows that $\Delta J_{l}$
reduce with increasing $\xi_{10}$ in a similar way as Fig. \ref{fig:individual-decay-dephasing}
(c) for $\gamma_{10}$ except that $\Delta J_{x}$ and $\Delta J_{y}$
are always larger than $\Delta J_{z}$. However, we should notice
that $\xi_{10}$ is in the range $\left[0,100\Gamma_{10}\right]$
while $\gamma_{10}$ is in $\left[0,20\Gamma_{10}\right]$. Thus,
$\gamma_{10}$ affects the steady-state superradiance much stronger
than $\xi_{10}$. Notice that $\gamma_{10}$ contributes not only to dephasing
but also affects the population directly.

\section{Angular Moment and Uncertainty\label{sec:angular-momentum-uncertainty}}

In this appendix, we present the formula to compute the angular moment
$\mathbf{J}=\sum_{i=x,y,z}J_{i}\mathbf{e}_{i}$ and its uncertainty
$\Delta\mathbf{J}=\sum_{i=x,y,z}\Delta J_{i}\mathbf{e}_{i}$. The
components $J_{x}$ and $J_{y}$ are expectation value of the angular
momentum operators $\sigma_{x}=\left(1/2\right)\left(\sigma_{+}+\sigma_{-}\right)$,
$\sigma_{y}=-\left(i/2\right)\left(\sigma_{+}-\sigma_{-}\right)$.
Thus, they can be calculated as $J_{x}=\left(1/2\right)\left(C_{10}+C_{01}\right)$
and $J_{y}=-\left(i/2\right)\left(C_{10}-C_{01}\right)$ with the
expectation value of the collective creation $\sigma^{+}$ and annihilation
operator $\sigma^{-}$: $C_{10}\equiv\left\langle \sigma^{+}\right\rangle =\sum_{l=0}^{N}C_{N}^{l}\left(N-l\right)\left\langle l,1,0,N-l-1\right\rangle ,$$C_{01}\equiv\left\langle \sigma^{-}\right\rangle =\sum_{l=0}^{N}C_{N}^{l}l\left\langle l-1,0,1,N-l\right\rangle $.
On the right side, the numbers from the left to right are $n_{11},n_{10},n_{01},n_{00}$.
The component $J_{z}\equiv\left(1/2\right)\left\langle \sigma_{z}\right\rangle $
can be calculated with the expectation value $\left\langle \sigma_{z}\right\rangle $
of the collective Pauli operator $\sigma_{z}=\sigma_{11}-\sigma_{00}$.
In return, this expectation value $\left\langle \sigma_{z}\right\rangle =P_{1}-P_{0}$
can be computed with the population of the upper level $P_{1}\equiv\left\langle \sigma_{11}\right\rangle $
and lower level $P_{0}\equiv\left\langle \sigma_{00}\right\rangle $,
respectively: $P_{1}=\sum_{l=0}^{N}C_{N}^{l}l\left\langle l,0,0,N-l\right\rangle ,$$P_{0}=\sum_{l=0}^{N}C_{N}^{l}\left(N-l\right)\left\langle l,0,0,N-l\right\rangle .$
The components $\Delta J_{i}$ of the uncertainty vector $\Delta\mathbf{J}$
can be calculated with $\Delta J_{i}=\sqrt{\left\langle j_{i}^{2}\right\rangle -J_{i}^{2}}$.
We can evaluate the expectation value of the square of the angular
momentum operators: 
\begin{align}
 & \left\langle j_{x}^{2}\right\rangle =\left(1/4\right)\sum_{l=0}^{N}C_{N}^{l}[l\left(l-1\right)\left\langle l-2,0,2,N-1\right\rangle \nonumber \\
 & +N\left\langle l,0,0,N-l\right\rangle +2l\left(N-l\right)\left\langle l-1,1,1,N-l-1\right\rangle \nonumber \\
 & +\left(N-l\right)\left(N-l-1\right)\left\langle l,2,0,N-l-2\right\rangle ],\label{eq:sigma_x} \\
 & \left\langle j_{y}^{2}\right\rangle =\left(-1/4\right)\sum_{l=0}^{N}C_{N}^{l}[l\left(l-1\right)\left\langle l-2,0,2,N-l\right\rangle \nonumber \\
 & -N\left\langle l,0,0,N-l\right\rangle -2l\left(N-l\right)\left\langle l-1,1,1,N-l-1\right\rangle \nonumber \\
 & +\left(N-l\right)\left(N-l-1\right)\left\langle l,2,0,N-l-2\right\rangle ],\label{eq:sigma_y} \\
 & \left\langle j_{z}^{2}\right\rangle =\left(1/4\right)\sum_{l=0}^{N}C_{N}^{l}\left(2l-N\right)^{2}\left\langle l,0,0,N-l\right\rangle .\label{eq:sigma_z}
\end{align}


\end{document}